\documentclass{article} 
\usepackage{iclr2027_conference,times}

\usepackage{amsmath,amsfonts,bm}

\def\1{\bm{1}}

\DeclareMathAlphabet{\mathsfit}{\encodingdefault}{\sfdefault}{m}{sl}
\SetMathAlphabet{\mathsfit}{bold}{\encodingdefault}{\sfdefault}{bx}{n}

\usepackage[hidelinks]{hyperref}
\usepackage{url}

\usepackage{graphicx}
\usepackage{booktabs}
\usepackage{amsmath,amssymb,amsfonts}
\usepackage{amsthm}
\usepackage{multirow}
\usepackage{array}
\usepackage[table]{xcolor}
\usepackage{algorithm}
\usepackage{algorithmic}
\usepackage{caption}
\usepackage{subcaption}
\usepackage{wrapstuff}
\usepackage{wrapfig}

\usepackage{microtype}          
\newcommand{\safewrap}[1][60pt]{\par}

\newtheoremstyle{fedbold}
  {3pt}{3pt}
  {\normalfont}{}
  {\bfseries}{.}
  {0.5em}{}
\theoremstyle{fedbold}
\newtheorem{theorem}{Theorem}

\newtheorem{proposition}[theorem]{Proposition}

\newtheorem{assumption}[theorem]{Assumption}
\theoremstyle{remark}

\title{FedPix: Learning What to Share with Less Leakage in Federated Intelligence}

\author{%
\bfseries
\mbox{Jingwei Zhao\textsuperscript{1,\textdagger}},
\mbox{Tiantong Sun\textsuperscript{2,\textdagger}},
\mbox{Yuanyuan Song\textsuperscript{1}},
\mbox{Jinhuan Guo\textsuperscript{1}},
\mbox{Zishuo Wei\textsuperscript{1}},
\mbox{Shengyu Huang\textsuperscript{1}},
\mbox{Ziyuan Ren\textsuperscript{1}},
\mbox{Wei Chen\textsuperscript{1}},
\mbox{Xiaojin Zhang\textsuperscript{1,*}}\\[0.5em]
\normalfont\small
\textsuperscript{1}Huazhong University of Science and Technology, Wuhan, China\\
\textsuperscript{2}George Mason University, Fairfax, VA, USA
}
\iclrpreprintcopy
\hypersetup{
  pdftitle={FedPix: Learning What to Share with Less Leakage in Federated Intelligence},
  pdfauthor={Jingwei Zhao, Tiantong Sun, Yuanyuan Song, Jinhuan Guo, Zishuo Wei, Shengyu Huang, Ziyuan Ren, Wei Chen, Xiaojin Zhang}
}

\begin{document}

\maketitle

\makeatletter
\begingroup
\renewcommand{\@makefntext}[1]{\raggedright\setlength{\parindent}{0pt}\noindent\@makefnmark\,#1}
\renewcommand{\thefootnote}{\textdagger}
\footnotetext[1]{Jingwei Zhao and Tiantong Sun contributed equally to this work and are co-first authors.}
\renewcommand{\thefootnote}{*}
\footnotetext[2]{Corresponding author: Xiaojin Zhang (\mbox{xiaojinzhang@hust.edu.cn}).}
\endgroup
\makeatother

\begin{abstract}
Existing federated intelligence paradigms primarily focus on how models are trained or aggregated, while overlooking a more fundamental question: \textbf{what information should leave the client?} Existing methods often share model updates that users cannot easily inspect, or data representations that may reveal sensitive information. Moreover, generating these representations can be computationally expensive for resource-constrained clients. We observe that image classification does not require all the visual details needed to reconstruct the original image. This suggests that clients could share less information while preserving task performance. However, learning what to preserve is challenging when the server cannot access raw client data. To address this challenge, we propose \textbf{FedPix}, which learns to share compact, visually inspectable images instead of model updates or intermediate features. The key idea is to use the shared representations themselves to learn what information should be preserved and what can be discarded. By balancing task utility and reconstruction resistance, FedPix enables the server to improve the representations without accessing raw client data, while keeping client-side computation lightweight. Extensive experiments on BloodMNIST, CIFAR-10, FashionMNIST, and MNIST demonstrate that FedPix achieves strong reconstruction resistance, with reconstruction PSNR as low as 6.87~dB and synthetic PSNR as low as 6.74~dB. Meanwhile, FedPix reduces client runtime by up to 96.0\% (from 1223~s to 49~s on CIFAR-10), while maintaining competitive classification accuracy of 99.18\% on MNIST, 89.97\% on FashionMNIST, 82.24\% on BloodMNIST, and 65.63\% on CIFAR-10.
\end{abstract}

\section{Introduction}
\label{sec:intro}

Existing federated intelligence paradigms mainly focus on how models are
trained or aggregated across clients, while overlooking a more
fundamental question: \emph{what information should leave the client in
the first place?} Personalized services---recommendation systems,
intelligent assistants, medical diagnosis---increasingly depend on
users' local data such as photos, medical images, and interaction
histories to capture user preferences and support downstream
tasks~\citep{ojokoh2025privacy}. Directly uploading raw data, however,
may expose users' private information. The core challenge is therefore
not only how to train models without centralizing raw data, but how to
design a principled \emph{data interface} that controls what
representation the client outputs to the server.

Existing federated paradigms provide only partial answers to this
question, and they can be understood as operating at three distinct
layers of the system stack.

\textbf{Model coordination layer.}
Federated learning (FL) retains raw data locally, yet
requires clients to perform iterative optimization and exchange
large, abstract, and user-unreadable model updates~\citep{mcmahan2017communication}. Under
heterogeneous client data, local optimization may also introduce
client drift before aggregation~\citep{karimireddy2020scaffold}.
This line of work addresses \emph{how models should be coordinated
across clients}, but does not specify what information should leave
each client---it simply transmits whatever the local optimizer produces.

\textbf{Representation transmission layer.}
Federated data condensation (FDC)
communicates human-readable synthetic samples, but the generation
process can be expensive on edge devices, and the resulting images
may still retain excessive similarity to the original
images~\citep{xiong2023feddm,holland2024collabdm,zhang2024fedsd2c}.
Split federated learning (SFL) reduces some client-side computation, but
transmits intermediate features that users find difficult to inspect or
compare against their original
data~\citep{thapa2022splitfed,flexsfl2025}.
These methods transmit representations rather than model parameters,
yet they optimize for either distribution fidelity (FDC) or training
efficiency (SFL), not for a controlled privacy--utility trade-off at
the data boundary.

\textbf{Data interface layer.}
Existing federated paradigms have not explicitly optimized a
privacy-controllable data interface that governs what information is
released from the client. Such an interface differs from conventional
compression or feature transmission in its optimization objective:
instead of minimizing storage cost, communication cost, or
reconstruction distortion alone, it explicitly learns which information
should be exposed through the client-to-server boundary according to
downstream utility and privacy requirements. Data-level differential
privacy provides formal protection through random perturbation, yet
often incurs a substantial loss of downstream
utility~\citep{dppix2019,dpvae2023,ldpslicing2026}. Conventional
compression and learned compression are primarily designed to preserve
visual fidelity rather than suppress reconstructable
information~\citep{wallace1992jpeg,sullivan2012overview,balle2018variational,balle2017end}.
Task-oriented data compression can, in principle, filter out both
privacy- and utility-irrelevant information; however, these methods
assume direct access to original data during optimization and mainly
target compression efficiency for downstream analytics, rather than
designing a privacy-aware interface for federated data
sharing~\citep{alvar2021scalable,azizian2024ppa}.

Different from existing task-oriented compression methods~\citep{alvar2021scalable,azizian2024ppa},
which optimize encoders with direct access to source data for
compression efficiency, FedPix studies representation interface
learning under federated constraints where raw client samples are
unavailable. The central objective is therefore not bitrate
optimization but controlling the information exposed through
client-to-server transmission while preserving downstream utility.

These limitations indicate that a practical federated data method
should satisfy three requirements:
\begin{itemize}
\setlength{\topsep}{4pt}\setlength{\partopsep}{0pt}%
\setlength{\itemsep}{1pt}\setlength{\parsep}{1pt}
\item[] \textbf{R1: User auditability.}
The representation transmitted by the client should remain in the same
readable form as the original data (e.g., image-to-image), enabling
direct user inspection.
\item[] \textbf{R2: Edge-device friendliness.}
The primary computational burden should be borne by the server whenever
possible, accommodating personal phones and laptops.
\item[] \textbf{R3: Reconstruction privacy with task utility.}
The representation should minimize information in the uploaded
representation that is useful for reconstructing the original input
while preserving the information required by the downstream task, thereby making the uploaded representation auditable by and passable through
user audit while still meeting the system's utility objectives.
\end{itemize}

We propose \textbf{FedPix}, a federated data interface that learns a
task-sufficient yet reconstruction-resistant representation interface before data leaves
the client. In this framework, each client applies a lightweight
representation encoder once to the original image and uploads a
compact, human-readable small-sized image together with its label. The
server upsamples the transmitted small-sized images, trains a
classifier on the resulting images, then improves the encoder, and
redistributes the updated encoder to participating clients.
Consequently, clients perform no backpropagation or iterative local
optimization. The encoder serves as the implementation module of this
interface, realizing the privacy--utility trade-off through a learnable
capacity bottleneck. FedPix provides an auditable transmission object
(R1), substantially reduces client-side computation (R2), and
explicitly optimizes the trade-off between reconstruction similarity
and classification utility (R3).

FedPix must improve the encoder without observing raw client
inputs. To this end, we develop an optimization strategy that operates
without access to client raw data: it first upsamples the uploaded
small-sized images to train a classifier, then obtains the classifier's
correct classification distribution over the small-sized images.
Treating the encoder and classifier as an integrated whole, for each
small-sized image that has been correctly classified, we upsample it
back to the original resolution and re-input it into the encoder for
a second encoding pass. The encoder is then optimized with a utility
objective---that the twice-encoded images remain correctly classified
by the classifier---and a privacy objective---that the dissimilarity
between the before-and-after encoded images is maximized. This
design encourages the encoder to correctly extract
classification-beneficial information while making the encoded images
as dissimilar as possible to their pre-encoding counterparts, thereby
reducing the feasibility of reconstruction attacks.

\begin{wrapstuff}[l,width=0.50\textwidth,type=figure,hsep=0.8em,vsep=6pt]
\centering
\includegraphics[width=\linewidth]{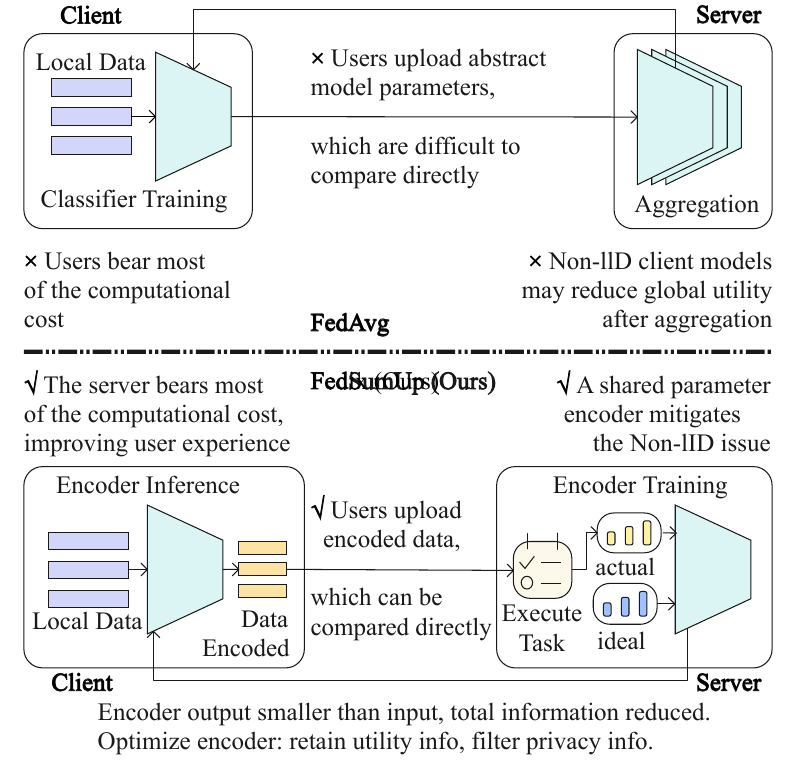}
\captionsetup{font=footnotesize,hypcap=false}
\caption{Traditional FL burdens clients with local training and communicates opaque model updates that users cannot audit, motivating the need for a lightweight, user-auditable data interface.}
\label{fig:schematic}
\end{wrapstuff}
A theoretical analysis characterizes the
task-preserving representation capacity region and identifies a
classifier-sensitive distortion mechanism, explaining when large visual
distortion can coexist with minimal classification loss.

Experiments on BloodMNIST, CIFAR-10, FashionMNIST, and MNIST evaluate
classification utility, reconstruction resistance, client computation,
communication cost, non-IID robustness, and membership inference.
FedPix reduces client runtime by up to 96.0\% while maintaining
competitive classification accuracy. More importantly, the
ACC--Synthetic-SSIM analysis explicitly evaluates the central
privacy--utility question: across the tested configurations, FedPix
lies on the empirical nondominated frontier on CIFAR-10 and
BloodMNIST, while the relative advantage is dataset-dependent on
FashionMNIST. These results support the intended mechanism: a learned
capacity bottleneck can remove visually reconstructable information
while retaining sufficient information for the downstream task.
\wrapstuffclear

\subsection{Contributions}

The contributions of this work are:
\begin{itemize}
\setlength{\topsep}{4pt}\setlength{\partopsep}{0pt}%
\setlength{\itemsep}{1pt}\setlength{\parsep}{1pt}

\item \textbf{An auditable federated data interface.}
We introduce a representation-sharing paradigm that treats
client-to-server transmission as an information selection interface.
Unlike conventional representation transmission, the proposed interface
learns a task-sufficient representation under federated constraints,
determining what information can be exposed to the server while
maintaining downstream utility. Clients upload compact, human-readable
images through a single encoder forward pass---not model updates, not
intermediate features, and not synthetic data---while all iterative
optimization is performed on the server.

\item \textbf{Raw-data-free representation evolution mechanism.}
We develop an upsample--recompress closed loop with
correctness-filtered utility preservation, enabling the server to train
a privacy-aware encoder without observing raw client inputs.

\item \textbf{A mechanism-level privacy--utility analysis (Appendix).}
We characterize the task-preserving representation capacity region,
identifying a free-representation regime where reconstruction distortion
can increase without increasing the Bayes cross-entropy.

\item \textbf{Comprehensive empirical evaluation.}
We evaluate utility, reconstruction resistance, auditability, client
cost, communication, non-IID behavior, and membership inference across
four datasets, including a direct ACC--Synthetic-SSIM frontier
analysis over multiple compression levels.

\end{itemize}

\section{Preliminaries}
\label{sec:prelim}

\subsection{Threat Model}
\label{subsec:threat}

We consider an \textbf{honest-but-curious server} that follows the
protocol but may infer private information from uploaded data
~\citep{shi2024dealing,hu2024source,shi2025unveiling}. Clients share an
intermediate representation $S=\Phi(X)$ and its label rather than raw
input $X$; $S$ may be synthetic data, a feature map, or a latent code
~\citep{xiong2023feddm,holland2024collabdm,erdogan2022unsplit,zhang2024fedsd2c}.
The server observes $S$. For FedPix, the server trains $\Phi$ and therefore knows its architecture and parameters. The optimization-based inversion attack below uses this encoder for gradient computation; Appendix~\ref{sec:appendix_adaptive_attack} additionally evaluates a white-box learned restoration attack.

\begin{wrapstuff}[l,width=0.46\textwidth,type=figure,hsep=0.8em,vsep=6pt]
\centering
\includegraphics[width=\linewidth]{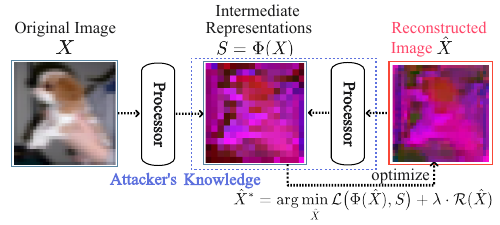}
\captionsetup{font=footnotesize,hypcap=false}
\caption{Model inversion attacks exploit intermediate representations to reconstruct original inputs; suppressing visually recoverable information in transmitted representations is the core defense strategy of FedPix.}
\label{fig:ir_process}
\end{wrapstuff}
We emphasize that FedPix targets empirical reconstruction resistance
rather than cryptographic privacy guarantees. The goal is to reduce
exploitable information exposed through client-to-server
representations under realistic inference attacks.
Labels are transmitted in the clear and are not protected by this interface.
The protocol also does not provide a formal bound on cumulative leakage
from repeated uploads across rounds.

We evaluate reconstructable leakage through the model inversion attack
of \citet{he2019model}:
\begin{equation}
\hat X^*=\arg\min_{\hat X}\mathcal L(\Phi(\hat X),S)
+\lambda\mathcal R_X(\hat X),
\label{eq:mia}
\end{equation}
where $\mathcal R_X$ imposes an image prior. FedPix does not claim to prevent all privacy attacks, but focuses on limiting visually recoverable information exposed through the client-server interface.

\section{Method}
\label{sec:method}

\subsection{Notation}
\label{subsec:notation}

We denote the original image and label at round $t$ by $D_t$ and $L_t$, where images have spatial size $S\times S$ (e.g., $32\times32$). The representation encoder $V_t$ maps an $S\times S$ input to a compact $y\times y$ output ($y<S$), producing the capacity-controlled representation $I_t=V_t(D_t)$ uploaded by the client. The server maintains a classifier $C_t$ and updates both modules iteratively. The joint training objective combines three terms: a utility consistency loss $\mathcal{L}_{\mathrm{ce}}$ (cross-entropy restricted to samples whose classifier prediction matches the ground-truth label, i.e.,~$\mathcal{I}_t^{\mathrm{correct}}=\{i\mid\hat{L}_{t,i}=L_{t,i}\}$), a structural dissimilarity loss $\mathcal{L}_{\mathrm{ssim}}$, and a pixel-level dissimilarity loss $\mathcal{L}_{\mathrm{psnr}}$, weighted by $\lambda$ and $\beta$ respectively. The representation dimension $y$ controls the physical capacity of the uploaded data.
\subsection{Complete Iterative Process}
\label{subsec:full_iteration}

Different from conventional compression systems that optimize encoders using accessible source samples, FedPix evolves the representation interface through server-side optimization because raw client data never participates in encoder training.

Given the original data pair $(D_t, L_t)$, the client encodes $D_t$ via the encoder: $I_t = V_t(D_t)$, where $V_t$ maps from $S \times S$ to $y \times y$ ($y < S$). The compressed representation $I_t$ and label $L_t$ are uploaded to the server.

The server then performs two sequential stages. \textbf{Stage~1 (Classifier Update):} The classifier $C_t$ is trained on upsampled representations to minimize cross-entropy: $C_{t+1} = \arg\min_{C}\; \text{CrossEntropy}(C(\text{UpSample}(I_t)), L_t)$.

\textbf{Stage~2 (Representation Interface Evolution):} The server updates $V_t$ through five steps.

\begin{wrapstuff}[l,width=0.45\textwidth,type=figure,hsep=0.7em,vsep=6pt]
\centering
\includegraphics[width=\linewidth]{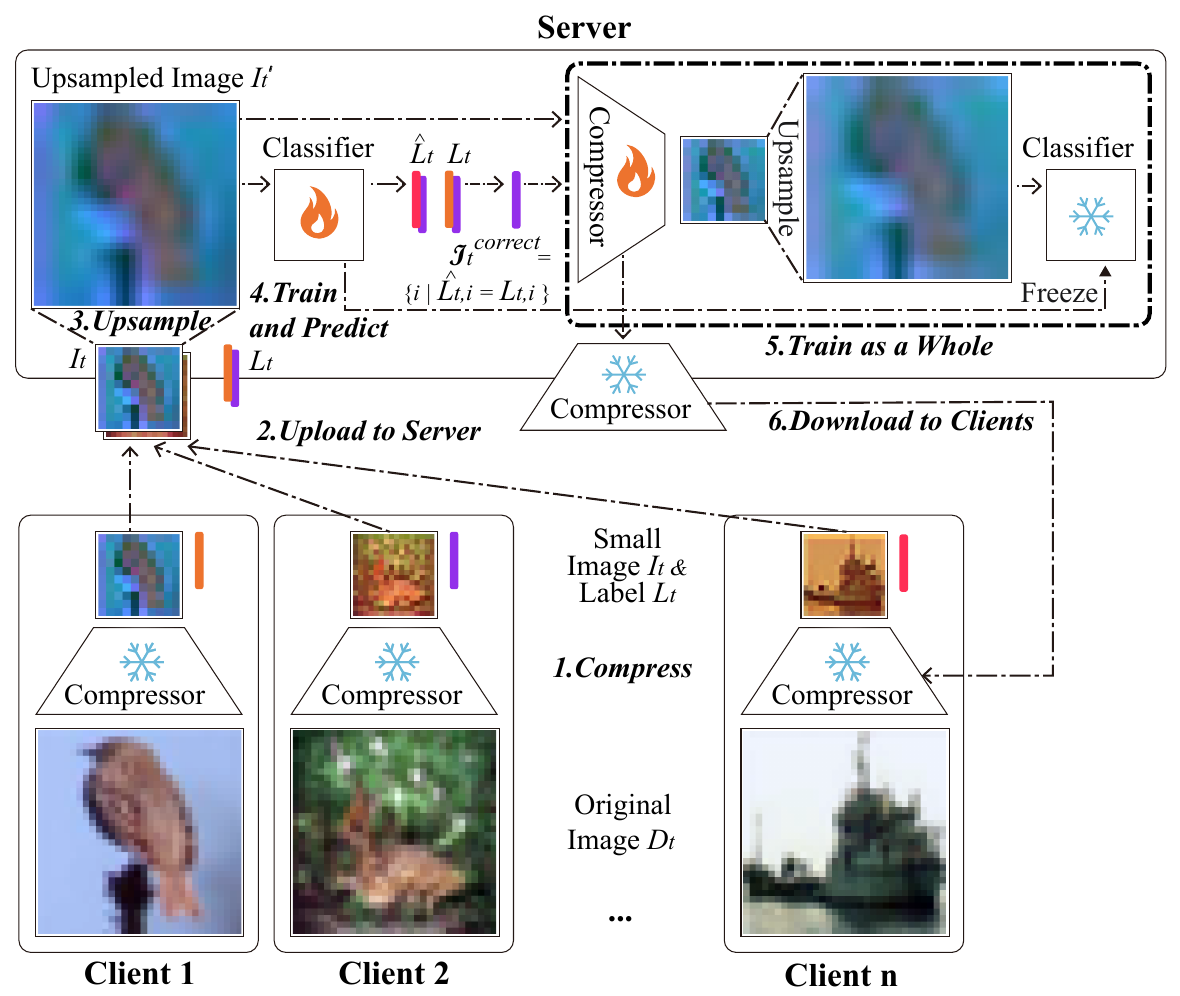}
\captionsetup{font=footnotesize,hypcap=false}
\caption{FedPix shifts computation to the server: clients only perform a single forward pass, while the server trains both the classifier and the encoder via an upsample--recompress closed loop without accessing raw data.}
\label{fig:framework_flowchart}
\end{wrapstuff}
\textit{Step~1 (Freeze classifier).} The parameters of $C_{t+1}$ are frozen to fix the decision boundary, so subsequent optimization reflects only the encoder's ability to preserve task-relevant information.

\textit{Step~2 (Identify consistent samples).} The frozen classifier predicts a hard label $\hat{L}_{t,i}$ by taking the class with the largest output on each upsampled representation. Correctly classified indices are recorded as $\mathcal{I}_t^{\text{correct}} = \{i \mid \hat{L}_{t,i} = L_{t,i}\}$. These samples participate in the encoder's utility-preservation training.

\textit{Step~3 (Upsample--recompress closed loop).} Since the server cannot access $D_t$, and the computation graph from $D_t$ to $I_t$ is broken at the client side, the server constructs a differentiable surrogate: $I'_t = \text{UpSample}(I_t)$, $I''_t = V_t(I'_t)$. This closed loop enables server-side optimization without raw data access, and trains $V_t$ to retain only the information sufficient for correct classification when recompressing the upsampled $I'_t$.

\textit{Step~4 (Joint loss).} For utility preservation, only consistent samples contribute to the cross-entropy: $\mathcal{L}_{\mathrm{ce}} = -\frac{1}{|\mathcal{I}_t^{\text{correct}}|} \sum_{i \in \mathcal{I}_t^{\text{correct}}} \log p(\hat{y}_i = L_{t,i} \mid I'''_t)$, where $I'''_t = \text{UpSample}(I''_t)$. For reconstruction suppression, we directly minimize $\mathcal{L}_{\mathrm{ssim}} = \text{SSIM}(\text{UpSample}(I''_t), I'_t)$ and $\mathcal{L}_{\mathrm{psnr}} = \text{PSNR}(\text{UpSample}(I''_t), I'_t)$. Minimizing these metrics does not provide formal differential privacy guarantees; instead, it explicitly discourages the preservation of visually recoverable structures commonly exploited by reconstruction attacks. The joint objective is:

\begin{equation*}
\mathcal{L}_{\mathrm{total}} = \lambda \cdot \mathcal{L}_{\mathrm{ce}} + \mathcal{L}_{\mathrm{ssim}} + \beta \cdot \mathcal{L}_{\mathrm{psnr}},
\end{equation*}

where $\lambda$ controls utility-preservation and $\beta$ controls reconstruction-suppression (SSIM term has fixed unit weight).

\textit{Step~5 (Gradient update).} $V_{t+1} = V_t - \eta \cdot \nabla_{V_t} \mathcal{L}_{\mathrm{total}}$.

\subsection{Inference Phase}
\label{subsec:inference}

After \(T\) training rounds, the framework produces a representation encoder \(V_{T-1}\) and a classifier \(C_T\). For a new sample \(D_{\text{new}}\), inference encodes it to \(I_{\text{new}}=V_{T-1}(D_{\text{new}})\), upsamples, and passes it to \(C_T\) for prediction \(\hat{y}_{\text{new}}\). Since \(V_{T-1}\) suppresses visually recoverable information, even if an attacker obtains \(I_{\text{new}}\) and attempts a model inversion attack, reconstructing the original input is difficult. Meanwhile, \(C_T\) is trained to classify accurately from such capacity-constrained representations, unifying reconstruction resistance and classification utility within a single framework.

\subsection{Theoretical Motivation}
\label{subsec:theory-motivation}

We provide a theoretical perspective on why capacity-constrained representations can improve reconstruction resistance without necessarily degrading task utility. Let $V$ denote the encoder mapping an image $D$ to a compact representation $I=V(D)$ of size $y\times y$ ($y<S$), and let $C$ be the server-side classifier. FedPix's objective is formalized as
$\min_{C,V}\; \mathcal L_{\mathrm{sim}}(V) \;\text{s.t.}\; \mathcal L_{\mathrm{cls}}(C,V) \leq \epsilon$,
with the Lagrangian relaxation $\min_{C,V}\; \lambda\mathcal L_{\mathrm{cls}}(C,V) + \mathcal L_{\mathrm{sim}}(V)$ (complete definitions in Appendix).

\begin{proposition}[Classifier-sensitive distortion]
\label{prop:sensitive-distortion-main}
Let $C_{t+1}$ be a classifier with $\kappa$-Lipschitz Jacobian, and let $\Delta_t$ denote the pixel distortion between the original image and its upsampled compressed representation. Then
$\left|\ell_{\mathrm{cls}}(C_{t+1}(I_t^{\mathrm{up}}),L_t) - \ell_{\mathrm{cls}}(C_{t+1}(D_t),L_t)\right|
\leq \sqrt{2}\bigl(\|\nabla_D C_{t+1}(D_t)\Delta_t\|_2 + \frac{\kappa}{2}\|\Delta_t\|_2^2\bigr).$
\end{proposition}

Proposition~\ref{prop:sensitive-distortion-main} separates pixel distortion from classifier-sensitive distortion: a compressed image can exhibit large visual deviation (low PSNR/SSIM) while incurring minimal classification loss, provided the distortion lies in classifier-insensitive directions. The proof is provided in the Appendix.

\begin{theorem}[Task-preserving capacity threshold]
Under a canonical task--nuisance decomposition of the image space, if the representation capacity $r_y$ exceeds the task-relevant dimension $d_{\mathrm{task}}$, there exists an encoder that discards nuisance information contributing to reconstruction while preserving optimal classification performance. When $r_y<d_{\mathrm{task}}$, no linear encoder can preserve all label-relevant information.
\end{theorem}

The full statement and proof (Appendix) identify a task-preserving capacity region: as long as $r_y\geq d_{\mathrm{task}}$, reconstruction error can increase without degrading classification performance. Together, these results provide a mechanism-level explanation for the empirical privacy--utility frontier.

\section{Experiments}\label{sec:experiments}

\subsection{Experimental Setup}
\label{subsec:experimental_setup}

\subsubsection{Datasets and Baseline Methods}
We evaluate FedPix on four standard datasets: BloodMNIST, CIFAR-10, FashionMNIST, and MNIST.

Experiments use 10 clients with full participation (join ratio = 1.0). Non-IID data distributions are simulated via a Dirichlet distribution ($\alpha=0.5$). Unless stated otherwise, early stopping is applied: training halts when accuracy improvement is below $1\%$ compared to the previous round. This reduces computational load on edge devices and avoids excessive communication rounds that degrade user experience. Experiment 1 (baseline comparison) is repeated 10 times and averaged; others are repeated 5 times and averaged.

Evaluation metrics include classification accuracy (ACC, measuring model utility), model inversion attack metrics (PSNR$\downarrow$, SSIM$\downarrow$, LPIPS$\uparrow$, measuring reconstruction resistance), client runtime (measuring computational efficiency), and total communication volume (measuring communication overhead).

PSNR, SSIM, and LPIPS are computed using the IQA-PyTorch toolbox~\citep{pyiqa}. Before computing PSNR and SSIM, all images are normalized to the range [0,1].

\begin{table*}[t]
\centering
\caption{FedPix achieves the best or second-best utility--privacy--efficiency across most metrics: highest ACC on 3 of 4 datasets with the lowest client runtime by a wide margin and competitive reconstruction resistance. ``---'' indicates an unreported or inapplicable metric under the evaluated protocol; it does not imply immunity to reconstruction.}
\label{tab:baseline_acc_cria}
\setlength{\tabcolsep}{2.5pt}
\renewcommand{\arraystretch}{1.05}
\footnotesize
\resizebox{\textwidth}{!}{%
\begin{tabular}{l|c|ccc|ccc|c}
\toprule[1.5pt]
\multicolumn{9}{c}{\textbf{MedMNIST-BloodMNIST}} \\
\midrule
Method & ACC$\uparrow$ (\%) & \multicolumn{3}{c}{Reconstructed Data} & \multicolumn{3}{c|}{Synthetic Data} & Time$\downarrow$ \\
 &  & PSNR$\downarrow$ & SSIM$\downarrow$ & LPIPS$\uparrow$ & PSNR$\downarrow$ & SSIM$\downarrow$ & LPIPS$\uparrow$ & (s) \\
\midrule
FedAvg (2017) & 74.35$\pm$11.39 & --- & --- & --- & --- & --- & --- & 241$\pm$108 \\
CollabDM (2024) & 60.25$\pm$2.94 & --- & --- & --- & 20.49$\pm$0.23 & 0.78$\pm$0.01 & 0.03$\pm$0.00 & 1830$\pm$170 \\
FedDM (2023) & 78.79$\pm$2.40 & --- & --- & --- & 33.99$\pm$0.92 & 0.98$\pm$0.00 & 0.00$\pm$0.00 & 754$\pm$308 \\
FedSD2C (2024) & 68.56$\pm$2.81 & 10.36$\pm$0.10 & 0.19$\pm$0.00 & 0.21$\pm$0.00 & 13.94$\pm$0.16 & 0.45$\pm$0.02 & 0.14$\pm$0.01 & 245$\pm$9 \\
SFL (2020) & 64.86$\pm$11.68 & 37.43$\pm$1.56 & 0.99$\pm$0.00 & 0.00$\pm$0.00 & --- & --- & --- & 103$\pm$43 \\
\rowcolor{gray!25}
\textbf{FedPix (Ours)} & \textbf{82.24$\pm$4.44} & \textbf{10.20$\pm$1.02} & \underline{0.34$\pm$0.08} & \textbf{0.26$\pm$0.03} & \textbf{8.84$\pm$1.34} & \textbf{0.32$\pm$0.10} & \textbf{0.26$\pm$0.02} & \textbf{13$\pm$5} \\
\midrule
\multicolumn{9}{c}{\textbf{CIFAR-10}} \\
\midrule
Method & ACC$\uparrow$ (\%) & \multicolumn{3}{c}{Reconstructed Data} & \multicolumn{3}{c|}{Synthetic Data} & Time$\downarrow$ \\
 &  & PSNR$\downarrow$ & SSIM$\downarrow$ & LPIPS$\uparrow$ & PSNR$\downarrow$ & SSIM$\downarrow$ & LPIPS$\uparrow$ & (s) \\
\midrule
FedAvg (2017) & 65.98$\pm$3.02 & --- & --- & --- & --- & --- & --- & 1223$\pm$370 \\
CollabDM (2024) & 31.86$\pm$1.19 & --- & --- & --- & 18.12$\pm$0.17 & 0.76$\pm$0.01 & 0.03$\pm$0.00 & 2570$\pm$164 \\
FedDM (2023) & 45.10$\pm$1.59 & --- & --- & --- & 30.14$\pm$0.54 & 0.96$\pm$0.00 & 0.01$\pm$0.00 & 2479$\pm$318 \\
FedSD2C (2024) & 44.15$\pm$0.79 & 8.83$\pm$0.10 & 0.12$\pm$0.00 & 0.22$\pm$0.00 & 12.27$\pm$0.10 & 0.30$\pm$0.01 & 0.22$\pm$0.01 & 493$\pm$16 \\
SFL (2020) & 42.12$\pm$8.76 & 38.81$\pm$0.43 & 1.00$\pm$0.00 & 0.00$\pm$0.00 & --- & --- & --- & 412$\pm$180 \\
\rowcolor{gray!25}
\textbf{FedPix (Ours)} & \underline{65.63$\pm$2.08} & \underline{10.94$\pm$1.12} & \underline{0.33$\pm$0.08} & \textbf{0.30$\pm$0.08} & \textbf{8.47$\pm$1.44} & \textbf{0.24$\pm$0.07} & \textbf{0.39$\pm$0.05} & \textbf{49$\pm$11} \\
\midrule
\multicolumn{9}{c}{\textbf{FashionMNIST}} \\
\midrule
Method & ACC$\uparrow$ (\%) & \multicolumn{3}{c}{Reconstructed Data} & \multicolumn{3}{c|}{Synthetic Data} & Time$\downarrow$ \\
 &  & PSNR$\downarrow$ & SSIM$\downarrow$ & LPIPS$\uparrow$ & PSNR$\downarrow$ & SSIM$\downarrow$ & LPIPS$\uparrow$ & (s) \\
\midrule
FedAvg (2017) & 87.12$\pm$4.52 & --- & --- & --- & --- & --- & --- & 1438$\pm$261 \\
CollabDM (2024) & 69.22$\pm$5.04 & --- & --- & --- & 17.24$\pm$0.16 & 0.69$\pm$0.01 & 0.03$\pm$0.00 & 3326$\pm$227 \\
FedDM (2023) & 81.06$\pm$0.91 & --- & --- & --- & 24.16$\pm$0.48 & 0.81$\pm$0.01 & 0.01$\pm$0.00 & 3263$\pm$584 \\
FedSD2C (2024) & 78.54$\pm$1.01 & 7.38$\pm$0.14 & 0.16$\pm$0.01 & 0.34$\pm$0.01 & 12.23$\pm$0.25 & 0.43$\pm$0.02 & 0.15$\pm$0.01 & 712$\pm$30 \\
SFL (2020) & 76.36$\pm$5.30 & 32.13$\pm$1.34 & 0.92$\pm$0.01 & 0.00$\pm$0.00 & --- & --- & --- & 398$\pm$172 \\
\rowcolor{gray!25}
\textbf{FedPix (Ours)} & \textbf{89.97$\pm$0.65} & \underline{8.19$\pm$0.98} & \underline{0.26$\pm$0.05} & \textbf{0.41$\pm$0.05} & \textbf{8.27$\pm$2.34} & \textbf{0.29$\pm$0.09} & \textbf{0.39$\pm$0.09} & \textbf{63$\pm$16} \\
\midrule
\multicolumn{9}{c}{\textbf{MNIST}} \\
\midrule
Method & ACC$\uparrow$ (\%) & \multicolumn{3}{c}{Reconstructed Data} & \multicolumn{3}{c|}{Synthetic Data} & Time$\downarrow$ \\
 &  & PSNR$\downarrow$ & SSIM$\downarrow$ & LPIPS$\uparrow$ & PSNR$\downarrow$ & SSIM$\downarrow$ & LPIPS$\uparrow$ & (s) \\
\midrule
FedAvg (2017) & 97.78$\pm$0.39 & --- & --- & --- & --- & --- & --- & 1107$\pm$320 \\
CollabDM (2024) & 83.53$\pm$1.76 & --- & --- & --- & 14.05$\pm$0.12 & 0.53$\pm$0.01 & 0.05$\pm$0.00 & 3471$\pm$231 \\
FedDM (2023) & 95.11$\pm$0.28 & --- & --- & --- & 25.94$\pm$0.58 & 0.73$\pm$0.01 & 0.00$\pm$0.00 & 2674$\pm$237 \\
FedSD2C (2024) & 95.31$\pm$0.73 & 5.82$\pm$0.06 & 0.18$\pm$0.01 & 0.36$\pm$0.00 & 14.33$\pm$0.18 & 0.47$\pm$0.01 & 0.06$\pm$0.01 & 814$\pm$61 \\
SFL (2020) & 96.09$\pm$2.14 & 37.90$\pm$0.94 & 0.88$\pm$0.02 & 0.00$\pm$0.00 & --- & --- & --- & 423$\pm$104 \\
\rowcolor{gray!25}
\textbf{FedPix (Ours)} & \textbf{99.18$\pm$0.21} & \underline{6.87$\pm$1.46} & \textbf{0.11$\pm$0.03} & \textbf{0.43$\pm$0.07} & \textbf{6.74$\pm$2.32} & \textbf{0.19$\pm$0.08} & \textbf{0.42$\pm$0.08} & \textbf{48$\pm$3} \\
\bottomrule[1.5pt]
\end{tabular}%
}
\end{table*}

\subsubsection{Comprehensive Baseline Comparison}

This experiment answers a core question: \textbf{Can FedPix simultaneously satisfy user auditability (R1), edge-device friendliness (R2), and the reconstruction-utility trade-off (R3)?} Results under the Non-IID setting ($\alpha=0.5$) are shown in Table~\ref{tab:baseline_acc_cria}.

For R2 (utility-efficiency), FedPix achieves the best or second-best ACC across all datasets while requiring the lowest client runtime by a wide margin. On BloodMNIST and FashionMNIST, FedPix attains the highest ACC (82.24\% and 89.97\%) with runtimes of only 13s and 63s---reductions of 94.6\% and 95.6\% versus FedAvg. On MNIST, FedPix achieves the highest ACC (99.18\%) while requiring merely 48s, a 95.7\% runtime reduction versus FedAvg. On CIFAR-10, FedPix's ACC (65.63\%) is on par with FedAvg (65.98\%) yet its runtime is merely 49s against FedAvg's 1223s (a 96.0\% reduction). Across all datasets, FedPix consistently ranks first in client efficiency while maintaining competitive or best accuracy.

For R3 (utility-reconstruction), FedPix attains best or second-best reconstructed data quality and ranks first across all synthetic data metrics---confirming strong resistance to model inversion attacks (MoIA). Notably, SFL achieves high-fidelity reconstruction (PSNR 32.13--38.81~dB), revealing the reconstruction risk of uncompressed feature transmission. FedAvg, which uploads model parameters rather than data representations, has no meaningful reconstruction or synthetic data metrics (all ``---''), as it operates at the model-coordination layer rather than the data-interface layer.

For R1, FedPix transmits compressed, human-readable image-like data supporting intuitive inspection---a capability absent in SFL and FedAvg which transmit opaque feature vectors or model updates. In contrast, existing baselines each sacrifice at least one requirement: CollabDM and FedDM incur heavy client-side computation, with runtimes up to over $140\times$ that of FedPix; FedSD2C compromises utility for reconstruction resistance; SFL lacks auditability and exposes raw information to reconstruction. In summary, FedPix satisfies R1--R3 simultaneously not by improving model aggregation, but by redesigning the data interface through which clients share information.

\subsection{Comparison with Differential Privacy and Lossy Compression Methods}

\begin{wrapstuff}[l,width=0.46\textwidth,type=figure,hsep=0.8em,vsep=6pt]
\centering
\includegraphics[width=0.72\linewidth]{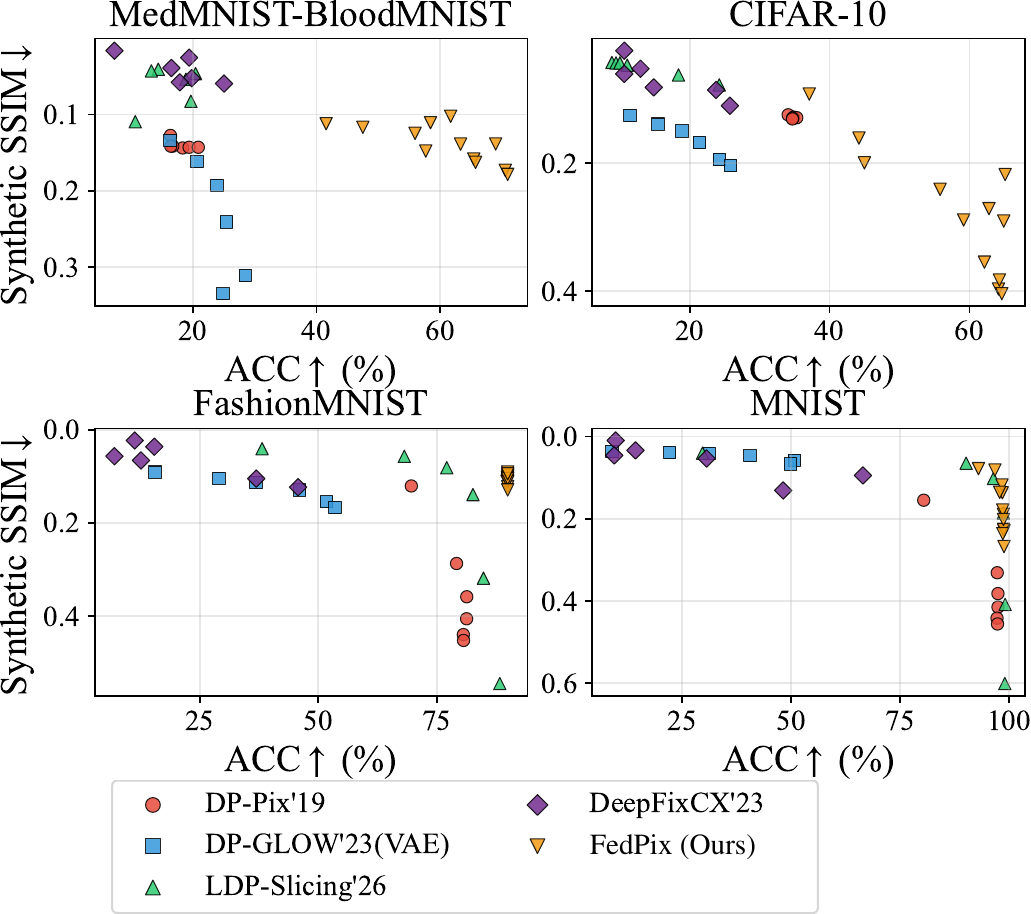}
\captionsetup{font=footnotesize,hypcap=false}
\caption{Empirical ACC--Synthetic-SSIM trade-offs across the evaluated configurations. Lower SSIM measures reduced direct image similarity, rather than a formal privacy guarantee.}
\label{fig:ssim_combined}
\end{wrapstuff}
We evaluate the ACC--Synthetic-SSIM trade-off across four datasets. Baselines include DP-Pix'19, DP-GLOW'23 (VAE), LDP-Slicing'26, and DeepFixCX'23. We sweep 6 noise levels for DP methods and 12 compressed image sizes for FedPix.

As shown in Figure~\ref{fig:ssim_combined}, FedPix lies on the empirical nondominated frontier on CIFAR-10 and BloodMNIST. Its relative advantage is configuration- and dataset-dependent on FashionMNIST and MNIST. The evaluated DP-based configurations often incur substantial accuracy costs. This comparison concerns observed accuracy and image similarity; it does not compare methods at matched formal privacy budgets, and FedPix provides no differential privacy guarantee.
\wrapstuffclear

\subsection{Impact of Compression Size on the Privacy--Utility Trade-off}

\begin{wrapstuff}[l,width=0.46\textwidth,type=figure,hsep=0.8em,vsep=6pt]
\centering
\includegraphics[width=\linewidth]{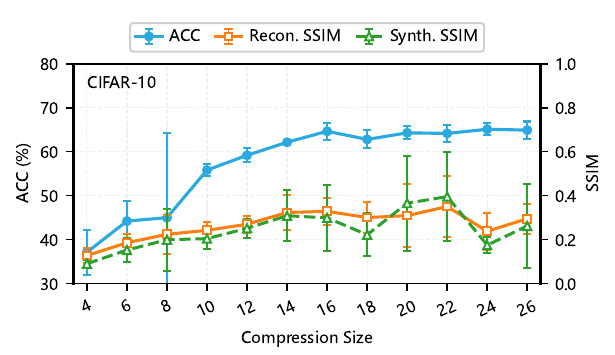}
\captionsetup{font=footnotesize,hypcap=false}
\caption{Smaller compression sizes ($y$) suppress reconstruction (lower Recon.\ SSIM) with minimal ACC loss, confirming that capacity constraints filter visually recoverable information while preserving task-relevant features.}
\label{fig:hyperparameter_cifar10}
\end{wrapstuff}
We sweep the compressed output size $y$ on CIFAR-10 to map the resulting privacy--utility trade-off; Figure~\ref{fig:hyperparameter_cifar10} reports the accuracy and reconstruction-similarity trends.

Figure~\ref{fig:hyperparameter_cifar10} shows the effect of varying the compressed output size on CIFAR-10. Increasing the size from 4 to 24 raises ACC from 37.10\% to 65.10\% with diminishing returns thereafter. Privacy is strongest at size 4 (Reconstruction SSIM 0.127, Synthetic SSIM 0.090) and weakest around size 22. Size~24 offers the best compromise---ACC 65.10\% with both SSIM values (0.238 and 0.175) lower than at sizes 20 and 22---confirming that the capacity bottleneck effectively controls the privacy--utility frontier.
\wrapstuffclear

\subsection{Ablation Study Analysis}

\begin{table}[t]
\centering
\caption{SSIM and PSNR losses are not always orthogonal to the utility objective: their effectiveness varies unpredictably across datasets and metrics, indicating that the interaction between reconstruction suppression and task preservation warrants further investigation.}
\label{tab:ablation}
\setlength{\tabcolsep}{3pt}
\newcommand{\ablstretch}{1.25}
\renewcommand{\arraystretch}{\ablstretch}
\resizebox{\columnwidth}{!}{%
\begin{tabular}{l|c|cc|cc}
\toprule[1.5pt]
\multicolumn{6}{c}{\textbf{BloodMNIST}} \\
\midrule
Variant & ACC (\%) $\uparrow$ & Syn.\ PSNR $\downarrow$ & Syn.\ SSIM $\downarrow$ & Recon.\ PSNR $\downarrow$ & Recon.\ SSIM $\downarrow$ \\
\midrule
Direct downsampling   & 86.58$\pm$3.09 & 25.22$\pm$0.21 & 0.8959$\pm$0.0023 & --- & --- \\
W/o anti-recon.\ loss & 76.48$\pm$13.05 & \textbf{6.73$\pm$1.45} & 0.2427$\pm$0.1369 & \textbf{8.18$\pm$1.00} & 0.2838$\pm$0.1412 \\
PSNR only             & \textbf{87.88$\pm$4.37} & 10.59$\pm$1.72 & 0.4990$\pm$0.1548 & 11.64$\pm$1.58 & 0.4648$\pm$0.1340 \\
SSIM only             & 75.47$\pm$10.28 & 6.89$\pm$1.19 & \textbf{0.2294$\pm$0.0927} & 8.72$\pm$1.06 & \textbf{0.2663$\pm$0.0935} \\
\rowcolor{gray!25} Complete FedPix & \cellcolor{gray!25} 81.75$\pm$4.75 & \cellcolor{gray!25} 9.32$\pm$1.54 & \cellcolor{gray!25} 0.3442$\pm$0.0932 & \cellcolor{gray!25} 10.97$\pm$1.50 & \cellcolor{gray!25} 0.3784$\pm$0.0843 \\
\midrule
\multicolumn{6}{c}{\textbf{FashionMNIST}} \\
\midrule
Variant & ACC (\%) $\uparrow$ & Syn.\ PSNR $\downarrow$ & Syn.\ SSIM $\downarrow$ & Recon.\ PSNR $\downarrow$ & Recon.\ SSIM $\downarrow$ \\
\midrule
Direct downsampling   & 89.72$\pm$0.15 & 31.65$\pm$0.23 & 0.9705$\pm$0.0017 & --- & --- \\
W/o anti-recon.\ loss & 89.72$\pm$0.15 & 9.50$\pm$0.94 & 0.1404$\pm$0.0187 & 10.04$\pm$0.33 & 0.1394$\pm$0.0194 \\
PSNR only             & 89.91$\pm$0.88 & 7.46$\pm$0.99 & 0.3323$\pm$0.0969 & 7.93$\pm$0.43 & 0.2806$\pm$0.0424 \\
SSIM only             & \textbf{89.99$\pm$0.50} & 6.45$\pm$0.93 & 0.2109$\pm$0.1046 & \textbf{6.92$\pm$1.10} & 0.2123$\pm$0.0931 \\
\rowcolor{gray!25} Complete FedPix & \cellcolor{gray!25} 89.70$\pm$0.11 & \cellcolor{gray!25} \textbf{5.59$\pm$0.37} & \cellcolor{gray!25} \textbf{0.1089$\pm$0.0386} & \cellcolor{gray!25} 8.98$\pm$0.56 & \cellcolor{gray!25} \textbf{0.1203$\pm$0.0250} \\
\midrule
\multicolumn{6}{c}{\textbf{CIFAR-10}} \\
\midrule
Variant & ACC (\%) $\uparrow$ & Syn.\ PSNR $\downarrow$ & Syn.\ SSIM $\downarrow$ & Recon.\ PSNR $\downarrow$ & Recon.\ SSIM $\downarrow$ \\
\midrule
Direct downsampling   & \textbf{68.66$\pm$1.48} & 23.49$\pm$0.22 & 0.8273$\pm$0.0043 & --- & --- \\
W/o anti-recon.\ loss & 65.68$\pm$2.40 & \textbf{8.17$\pm$2.51} & \textbf{0.3163$\pm$0.2202} & 9.74$\pm$2.61 & \textbf{0.3633$\pm$0.2096} \\
PSNR only             & 67.67$\pm$1.88 & 8.85$\pm$0.45 & 0.3297$\pm$0.0713 & 10.04$\pm$0.31 & 0.3660$\pm$0.0857 \\
SSIM only             & 65.49$\pm$2.01 & 8.52$\pm$1.78 & 0.3225$\pm$0.1357 & \textbf{9.55$\pm$1.88} & 0.3645$\pm$0.1325 \\
\rowcolor{gray!25} Complete FedPix & \cellcolor{gray!25} 66.66$\pm$0.44 & \cellcolor{gray!25} 9.16$\pm$0.76 & \cellcolor{gray!25} 0.3751$\pm$0.0825 & \cellcolor{gray!25} 10.34$\pm$1.07 & \cellcolor{gray!25} 0.4166$\pm$0.0704 \\
\midrule
\multicolumn{6}{c}{\textbf{MNIST}} \\
\midrule
Variant & ACC (\%) $\uparrow$ & Syn.\ PSNR $\downarrow$ & Syn.\ SSIM $\downarrow$ & Recon.\ PSNR $\downarrow$ & Recon.\ SSIM $\downarrow$ \\
\midrule
Direct downsampling   & 99.15$\pm$0.10 & 22.13$\pm$0.10 & 0.8481$\pm$0.0017 & --- & --- \\
W/o anti-recon.\ loss & \textbf{99.19$\pm$0.12} & 6.41$\pm$3.61 & 0.2120$\pm$0.0798 & 6.74$\pm$3.39 & 0.1746$\pm$0.0686 \\
PSNR only             & 99.03$\pm$0.21 & \textbf{6.31$\pm$1.56} & \textbf{0.1288$\pm$0.0210} & \textbf{6.70$\pm$1.20} & \textbf{0.0894$\pm$0.0250} \\
SSIM only             & 99.09$\pm$0.22 & 8.66$\pm$2.52 & 0.2286$\pm$0.1006 & 7.66$\pm$1.33 & 0.1691$\pm$0.0593 \\
\rowcolor{gray!25} Complete FedPix & \cellcolor{gray!25} 98.97$\pm$0.43 & \cellcolor{gray!25} 8.16$\pm$3.20 & \cellcolor{gray!25} 0.2267$\pm$0.0659 & \cellcolor{gray!25} 7.19$\pm$1.97 & \cellcolor{gray!25} 0.0903$\pm$0.0187 \\
\bottomrule[1.5pt]
\end{tabular}%
}
\end{table}
We conduct an ablation comparing five variants across four datasets (Table~\ref{tab:ablation}): (1)~\textbf{Direct downsampling}: bicubic downsampling replaces the learnable encoder; (2)~\textbf{W/o anti-recon.\ loss}: both SSIM and PSNR terms are removed ($\mathcal{L}_{\mathrm{ssim}}{=}0$, $\mathcal{L}_{\mathrm{psnr}}{=}0$), retaining only $\mathcal{L}_{\mathrm{ce}}$; (3)~\textbf{PSNR only}: only the PSNR loss is active ($\mathcal{L}_{\mathrm{ssim}}{=}0$); (4)~\textbf{SSIM only}: only the SSIM loss is active ($\mathcal{L}_{\mathrm{psnr}}{=}0$); (5)~\textbf{Complete FedPix}: both SSIM and PSNR losses are active.

Direct downsampling yields high synthetic PSNR/SSIM (e.g., 22--32 dB, 0.83--0.97), confirming that naive spatial reduction provides negligible privacy protection. Removing both anti-reconstruction losses achieves the lowest synthetic PSNR on BloodMNIST and CIFAR-10, and the lowest synthetic SSIM on CIFAR-10, but it does not consistently dominate the other variants. Its mean accuracy is lower than that of the complete model on these two datasets.

The independent activation of SSIM and PSNR reveals complementary roles. On MNIST, PSNR-only dominates nearly all privacy metrics (Syn.\ SSIM 0.13, Recon.\ PSNR 6.70, Recon.\ SSIM 0.09) while maintaining competitive accuracy (99.03\%). On FashionMNIST, the complete model achieves the best synthetic privacy (Syn.\ PSNR 5.59, Syn.\ SSIM 0.11), while SSIM-only excels at reconstruction resistance (Recon.\ PSNR 6.92). On BloodMNIST, SSIM-only achieves the best reconstruction SSIM (0.27), and PSNR-only achieves the highest accuracy (87.88\%). On CIFAR-10, the ``w/o'' variant leads in all synthetic and reconstruction SSIM, but SSIM-only achieves the best reconstruction PSNR (9.55).

The results reveal that the interaction between anti-reconstruction losses and the utility objective is \textbf{not consistently orthogonal}. On MNIST, PSNR-only dominates nearly all privacy metrics (Syn.\ SSIM 0.1288, Recon.\ SSIM 0.0894) while the complete model has higher Syn.\ SSIM (0.2267) and slightly higher Recon.\ SSIM (0.0903), suggesting interference between the loss terms. On CIFAR-10 and BloodMNIST, the variant without anti-reconstruction loss achieves the best privacy in several metrics, yet the complete model incurs a visible privacy cost (higher SSIM/PSNR) without a commensurate accuracy gain. On FashionMNIST, the complete model achieves the best synthetic privacy, but SSIM-only excels at reconstruction resistance. The ``best'' variant differs across datasets and metrics without a clear pattern. This instability---where adding a privacy-targeted loss sometimes \textit{worsens} privacy---indicates that SSIM/PSNR minimization is not always aligned with utility preservation in gradient space. Understanding and resolving this misalignment, perhaps through adaptive loss weighting or gradient projection, is a direction for future work.

\subsection{Utility--Privacy Pareto Analysis via MIA}
\label{sec:mia_pareto}

\begin{wrapstuff}[l,width=0.46\textwidth,type=figure,hsep=0.8em,vsep=6pt]
\centering
\includegraphics[width=0.82\linewidth]{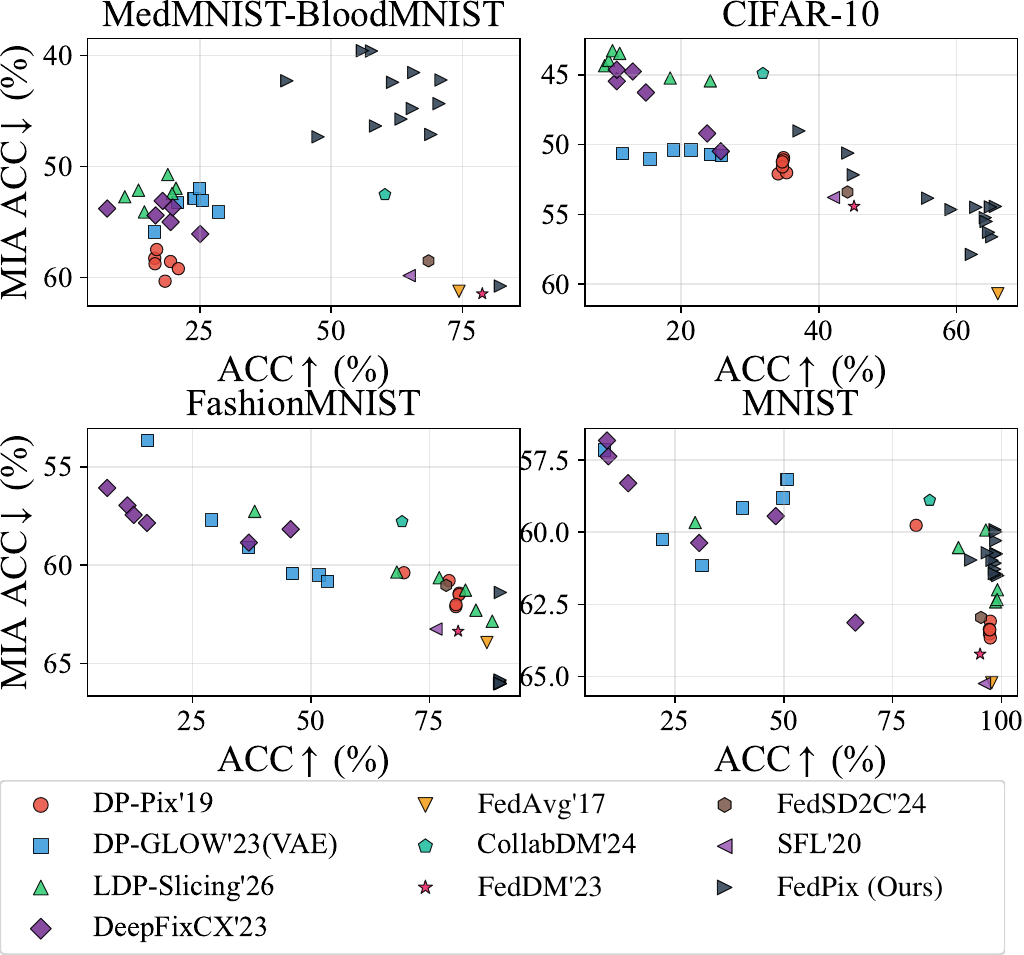}
\captionsetup{font=footnotesize,hypcap=false}
\caption{Classification accuracy versus membership-inference attack accuracy for the evaluated configurations. Attack accuracy near chance measures the success of this attack, rather than a general privacy guarantee.}
\label{fig:mia_combined}
\end{wrapstuff}

\textbf{MIA Setup.} Following \citet{shokri2017membership}, we adopt the standard shadow-model-based MIA protocol. The adversary (the honest-but-curious server) trains shadow models on auxiliary data, then trains a binary attack classifier on the shadow models' outputs to distinguish member from non-member samples. MIA ACC close to 50\% indicates the adversary performs no better than random guessing.

Figure~\ref{fig:mia_combined} compares FedPix against DP methods (DP-Pix'19, DP-GLOW'23, LDP-Slicing'26, DeepFixCX'23) and FL baselines (FedAvg'17, CollabDM'24, FedDM'23, FedSD2C'24, SFL'20) on the ACC--MIA ACC plane.

On CIFAR-10 and BloodMNIST, FedPix occupies the Pareto-optimal region: its 12 compressed-size configurations span 37--65\% ACC with MIA ACC of 49--58\% on CIFAR-10, substantially outperforming DP methods (DP-Pix peaks at 35\% ACC). On MNIST and FashionMNIST, FedPix achieves the best accuracy (99.18\% and 89.97\%) with competitive MIA ACC.

The evaluated DP configurations exhibit a substantial utility cost on CIFAR-10, with classification accuracy below 36\%. FedPix provides competitive empirical ACC--MIA trade-offs in the tested settings. However, near-chance attack accuracy does not establish absence of membership leakage, and these experiments do not compare matched differential privacy budgets. The membership-inference results complement the reconstruction evaluation; neither establishes a formal privacy guarantee for FedPix.
\wrapstuffclear

\section{Conclusion}\label{sec:conclusion}
We propose FedPix, a Federated Data Interface where clients upload capacity-constrained representations and the server optimizes both classifier and encoder via an upsample--recompress closed loop. FedPix achieves user auditability, edge-device friendliness (client runtime reduced up to 96\%), and strong reconstruction resistance with competitive accuracy, while remaining inherently robust to non-IID data.

\textbf{Limitations.} First, the capacity bottleneck (an $8\times8$ spatial compression) is fixed globally rather than adapted per task or dataset; our theoretical capacity region points toward adaptive mechanisms but does not yet realize them. Second, the evaluation focuses on image classification with a single encoder architecture, and the interface has not been validated on other modalities or tasks. Third, the on-device deployment benchmarks FP32 CPU inference only; INT8 quantization and NPU acceleration, which could further shrink the client footprint, remain unexplored. Finally, reconstruction resistance is evaluated against strong but specific attack families (model inversion and white-box restoration); future stronger attacks, e.g., diffusion-based reconstruction, could narrow the observed privacy margins.

\textbf{Future work.} Beyond addressing the above, we plan to (i) develop adaptive capacity mechanisms that negotiate the compression ratio per task and per user, (ii) extend the interface to non-image modalities and downstream tasks, (iii) deploy quantized, NPU-accelerated client encoders, and (iv) integrate the data interface with downstream federated aggregation and user-facing audit tooling, allowing participants to inspect exactly what leaves their devices.


\subsection*{Ethics statement}

This work studies representation sharing in federated intelligence and uses
only publicly available benchmark datasets (BloodMNIST, CIFAR-10,
FashionMNIST, and MNIST); it does not involve human subjects, personal data
collection, or model release. The threat model considered throughout the
paper is an honest-but-curious server, and the reconstruction and membership
inference attacks reported here are used solely to quantify the leakage of
the proposed data interface, not to attack deployed systems. We note that the
representation interface we propose reduces, but does not cryptographically
eliminate, the information recoverable from client-to-server transmissions,
and we state this limitation explicitly in the threat model
(Section~\ref{subsec:threat}).

\subsection*{Reproducibility statement}

We describe the experimental protocol in
Section~\ref{sec:experiments} and the appendix. The encoder architecture (including layer counts, hidden channels, pooling
configuration, and output format) and classifier are described in
Appendix~\ref{sec:method_appendix}; the
on-device deployment configuration and its end-to-end benchmark are reported
in Appendix~\ref{sec:appendix_ondevice}. Dataset preprocessing, client
counts, the Dirichlet heterogeneity parameter, early-stopping rule, and the
number of averaged repetitions are stated in
Section~\ref{subsec:experimental_setup}. The definitions and toolbox used for
all reconstruction metrics (PSNR, SSIM, LPIPS) are provided in
Appendix~\ref{sec:appendix_metrics}, and the attack settings---both the model
inversion attack of Section~\ref{subsec:threat} and the white-box adaptive
attack---are specified in Appendices~\ref{sec:method_appendix}~and
\ref{sec:appendix_adaptive_attack}. All theoretical claims are proven in
Appendix~\ref{app:theory-proofs}: the classifier-sensitive distortion bound
(Proposition~\ref{prop:sensitive-distortion}) and the task-preserving
capacity threshold (Theorem~\ref{thm:capacity-threshold}) include the
assumptions and proofs. These arguments characterize an idealized model
and do not establish privacy guarantees for the implemented nonlinear encoder.

\bibliography{aaai2027}
\bibliographystyle{iclr2027_conference}

\clearpage
\appendix

\section{Reconstruction Resistance Metrics}
\label{sec:appendix_metrics}

We quantify representation leakage from two complementary perspectives. First, \textbf{Synthetic--to--Original} measures the direct similarity between transmitted synthetic or compressed representations and the original data. Second, \textbf{Reconstructed--to--Original} measures the similarity between model-inversion-reconstructed samples and the original data. Following prior reconstruction and perceptual similarity studies~\citep{sun2023privacy, zhang2018unreasonable}, we use PSNR, SSIM, and LPIPS as evaluation metrics. Higher PSNR/SSIM and lower LPIPS indicate greater similarity to the original data and thus greater reconstruction leakage.

These metrics require access to original data and are therefore used only from the researcher's evaluation perspective. The HbC server in the actual protocol does not directly access raw data; it only observes uploaded intermediate representations.

We next provide a detailed rationale for each design choice, explaining how it concretely addresses the three requirements R1--R3.

\section{Detailed Design Rationale of FedPix}
\label{sec:method_appendix}

This section provides a detailed discussion of each design choice in FedPix and how it addresses the three requirements identified in Introduction: \textbf{R1 (User Auditability)}, \textbf{R2 (Edge-device Friendliness)}, and \textbf{R3 (Balancing Dissimilarity, Non-Reconstructability, and Utility)}.

\subsection{Why Train a Unified Encoder on the Server? (Addressing R2 and R3)}

A naive approach that trains encoders locally on each client faces two fundamental limitations. Incompatible non-IID encoders learned by separate clients cannot be reasonably aggregated, and running training procedures on edge hardware introduces substantial computational overhead.
FedPix adopts a \textbf{single unified encoder} trained centrally on the server and distributed every communication round, which fully removes client-side optimization workload. Clients only execute a single forward pass, so local objective divergence, client drift, and extra corrective stabilization mechanisms become unnecessary.
Regardless of data heterogeneity, the server's classifier receives inputs within a homogeneous encoding domain. This circumvents the pervasive non-IID challenge in conventional parameter-aggregation federated learning.
This design supports \textbf{R2 (Edge-device Friendliness)} by limiting client computation, and addresses \textbf{R3 (Balancing Dissimilarity, Non-Reconstructability, and Utility)}. Consistent encoding stabilizes decision boundaries and prevents utility loss caused by mismatched client-specific encoders. We stress that this centralized design is intentional: FedPix targets the data-interface layer (what information leaves the client), instead of the model-coordination layer for local model aggregation.

\subsection{Why the ``Upsample-Recompress'' Closed Loop? (Addressing R2)}

The server cannot access raw client data, which prevents end-to-end encoder training using paired original and compressed samples.
FedPix exploits the property that upsampling introduces no new statistical information. The server upsamples uploaded compressed representation $I_t$ to $I_t'$, then recompresses it via $V_t$ to obtain $I''_t = V_t(I_t')$. This closed-loop construction simulates client-side compression without raw inputs and enables gradient-based optimization hosted entirely on the server.
All training procedures run server-side, and clients only perform one forward pass for compression, with no backpropagation or iterative local optimization required.
This design directly supports \textbf{R2 (Edge-device Friendliness)}, shifting all optimization burden away from resource-limited edge devices.

\subsection{Why Minimize SSIM and PSNR for Reconstruction Suppression? (Addressing R3)}

Model inversion attacks take compressed images as initialization to reconstruct original inputs. Reconstruction difficulty rises when compressed outputs exhibit larger structural and pixel-level differences relative to raw data.
FedPix directly minimizes SSIM and PSNR measured between the upsampled re-encoded output $\text{UpSample}(I''_t)$ and the upsampled input $I'_t$. Since SSIM and PSNR both increase with similarity, minimizing these loss terms pushes the encoder to reduce the similarity between its output and the input, enlarging overall divergence.
SSIM captures structural discrepancies in luminance, contrast and spatial layout, while PSNR reflects pixel-level intensity deviations. The two metrics deliver complementary signals to obstruct visual reconstruction.
By actively maximizing divergence between compressed representations and upsampled data, FedPix suppresses visually recoverable information and mitigates inversion risks, which directly addresses \textbf{R3 (Balancing Dissimilarity, Non-Reconstructability, and Utility)}.

\subsection{Why Use Classifier Correctness as a Utility Filter? (Addressing R3)}

Pure minimization of SSIM and PSNR will indiscriminately erase all signal, including task-relevant information critical for classification.
FedPix introduces an auxiliary classifier $C$ trained and frozen on the server. The classifier generates hard-label predictions on upsampled compressed data; only samples whose predictions match ground-truth labels ($\mathcal{I}_t^{\mathrm{correct}}$) contribute to the utility consistency loss $\mathcal{L}_{\mathrm{ce}}$.
This constraint guides the encoder to preserve information shaping decision boundaries while suppressing visually recoverable content under bandwidth limits.
The loss $\mathcal{L}_{\mathrm{ce}}$ sustains task utility, while SSIM and PSNR minimization enforces dissimilarity and non-reconstructability. Their combination achieves a favorable reconstruction–utility trade-off and fulfills the demands of \textbf{R3}.

\subsection{Why Compressed Representations Preserve the Original Data Format? (Addressing R1)}

Split learning transmits abstract feature vectors that lack intuitive interpretability for end users.
Different from such paradigms, FedPix's encoder $V_t$ outputs a downsampled image $I_t$ that retains the original visual format (e.g., RGB channels, spatial layout) at reduced resolution $y \times y$.
The compressed representation maintains an intuitive correspondence with raw data. Users can compare uploaded content against original samples to audit privacy risks: one may identify coarse semantic content, although visual inspection alone cannot certify that fine-grained personal details are unrecoverable.
This transparency satisfies \textbf{R1 (User Auditability)}, enabling non-expert users to understand, audit, and consent to shared information — a benefit unavailable for opaque feature vectors or encrypted model parameters.

\subsection{Why Use an Additive Loss with SSIM and PSNR? (Addressing R3)}

Single dissimilarity metrics struggle to jointly control structural and pixel-level divergence, motivating an additive composite loss formulation.
FedPix adopts $\lambda \mathcal{L}_{\mathrm{ce}} + \mathcal{L}_{\mathrm{ssim}} + \beta \mathcal{L}_{\mathrm{psnr}}$. SSIM is naturally bounded ($[-1, 1]$) and contributes stable gradients, so its weight is fixed to unity. PSNR operates on a wider numerical scale (typically 5--40 dB), so $\beta$ is empirically selected to balance the gradient magnitudes contributed by the two dissimilarity terms. The tunable $\lambda$ separately regulates classification utility.
The additive form decouples structural and pixel-wise suppression. SSIM and PSNR provide different gradient signals, but their combination does not uniformly outperform isolated terms; the ablation in Table~\ref{tab:ablation} shows dataset-dependent interactions.
This loss construction delivers balanced, interpretable adjustment over the reconstruction–utility frontier and directly addresses \textbf{R3 (Balancing Dissimilarity, Non-Reconstructability, and Utility)}.

\subsection{Pseudocode of FedPix}

\begin{algorithm}[H]
\caption{The server optimizes the encoder without raw data by freezing the classifier, identifying correctly classified samples, and minimizing a joint loss over an upsample--recompress closed loop.}
\label{alg:framework}
\begin{algorithmic}[1]
\REQUIRE Client datasets $\{(D_k, L_k)\}$, initial encoder $V_0$, initial classifier $C_0$, number of rounds $T$, learning rate $\eta$
\ENSURE Inference pair $V_{T-1}, C_T$ (the classifier is trained on representations from $V_{T-1}$)
\FOR{$t = 0$ to $T-1$}
    \STATE \textbf{// Client}
    \STATE $I_t = V_t(D_t)$ \COMMENT{$V$ inputs $S \times S$, outputs $y \times y$}
    \STATE Upload $(I_t, L_t)$ to server
    \STATE \textbf{// Server Stage 1: Classifier Learning}
    \STATE $I_t^{\text{up}} = \text{UpSample}(I_t)$ \COMMENT{Upsample to $S \times S$}
    \STATE $C_{t+1} = \text{train}(C_t, I_t^{\text{up}}, L_t)$
    \COMMENT{Use all labels}
    \STATE \textbf{// Server Stage 2: Encoder Self-Evolution}
    \STATE Freeze $C_{t+1}$
    \STATE $\hat{L}_{t,i} = \arg\max_c [C_{t+1}(I_{t,i}^{\text{up}})]_c$ \COMMENT{Compute hard labels}
    \STATE $\mathcal{I}_t^{\text{correct}} = \{i \mid \hat{L}_{t,i} = L_{t,i}\}$
    \COMMENT{Correctly classified indices}
    \STATE $I'_t = \text{UpSample}(I_t)$ \COMMENT{Upsample to $S \times S$}
    \STATE $I''_t = V_t(I'_t)$ \COMMENT{$V$ re-encodes to $y \times y$}
    \STATE $I'''_t= \text{UpSample}(I''_t)$
    \COMMENT{Upsample to $S \times S$ for $C$}
    \STATE $\mathcal{L}_{\text{ce}} = \text{CE}_{\mathcal{I}_t^{\text{correct}}}(C_{t+1}(I'''_t), L_t)$
    \STATE $\mathcal{L}_{\text{ssim}} = \text{SSIM}(\text{UpSample}(I''_t), I'_t)$
    \STATE $\mathcal{L}_{\text{psnr}} = \text{PSNR}(\text{UpSample}(I''_t), I'_t)$
    \STATE $\mathcal{L}_{\text{total}} = \lambda \cdot \mathcal{L}_{\text{ce}} + \mathcal{L}_{\text{ssim}} + \beta \cdot \mathcal{L}_{\text{psnr}}$
    \STATE $V_{t+1} = V_t - \eta \cdot \nabla \mathcal{L}_{\text{total}}$
    \STATE Send $V_{t+1}$ back to client
\ENDFOR
\RETURN $V_{T-1}, C_{T}$
\COMMENT{Used jointly for inference: new samples first encoded by $V_{T-1}$, then classified by $C_T$}
\end{algorithmic}
\end{algorithm}

\subsection{Model Inversion Attack Settings}

The model inversion attack uses an encoder to reverse-reconstruct original images from synthetic data via gradient descent. The reconstruction process uses the Adam optimizer, a learning rate of 0.1, 500 iterations, and a total variation regularization weight of 0.001. After reconstruction, PSNR, SSIM, and LPIPS are computed between the reconstructed and original images (see Section~\ref{sec:appendix_metrics} for metric definitions).

The reconstruction attack evaluation procedure is as follows: obtain the aggregated synthetic data and their corresponding original data indices from the server side, uniformly select samples by category (10 per category), and compute the metrics between the reconstructed data and the original data.

\subsection{FedPix Implementation Details}

The representation encoder uses a convolutional architecture with an enhanced ConvPoolBlock. By default, it has five $3\times3$ convolutional layers, each followed by GroupNorm and ReLU, ending with AdaptiveAvgPool2d to downsample to the target spatial size. The hidden layers have 128 channels, and the output is a 3-channel RGB image produced by a $1\times1$ convolution with $\tanh$ activation (pixel range $[-1,1]$). A Squeeze-and-Excitation channel attention mechanism recalibrates feature maps via global average pooling and a two-layer fully connected network. For CIFAR-10, the encoder compresses a $3\times32\times32$ input to $3\times8\times8$ ($192$ values), achieving a $16\times$ compression ratio (from $3072$ values). When the compressed image size is set to 24, the output is $3\times24\times24$, giving a $1.78\times$ compression ratio. The classifier follows the ConvNet architecture used in FedAvg.

\section{Hyperparameter Analysis}
\label{sec:appendix_hyperparam}

\subsection{Impact of Utility Ratio \texorpdfstring{$\lambda$}{lambda} on the Privacy--Utility Trade-off}

This experiment answers one question --- how does the utility ratio $\lambda$ effectively control the preference direction of the privacy-utility trade-off as a trade-off coefficient in the joint optimization objective?

\begin{wrapfigure}{l}{0.50\textwidth}
\centering
\includegraphics[width=\linewidth]{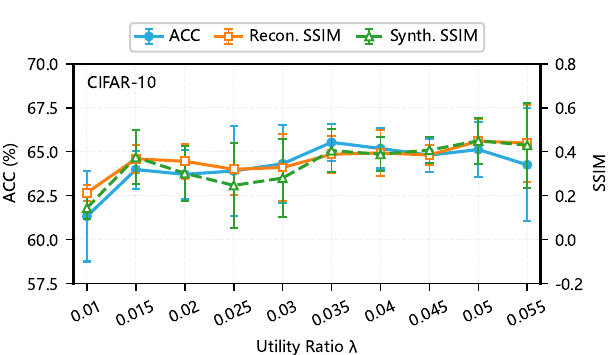}
\captionsetup{font=footnotesize,hypcap=false}
\caption{Increasing $\lambda$ improves ACC up to a saturation point ($\lambda \approx 0.035$) beyond which information leakage (SSIM) rises without commensurate utility gain, indicating an optimal operating region.}
\label{fig:utility_ratio_ablation_ssim}
\end{wrapfigure}
The experiment was conducted by scanning different values of $\lambda$ under the same federated learning setup. The results (Fig.~\ref{fig:utility_ratio_ablation_ssim}) show that when $\lambda$ is too small (e.g., 0.01), the compressor prioritizes privacy over utility, leading to the lowest classification accuracy (61.32\%) but the strongest privacy protection, with the reconstruction SSIM and synthetic SSIM reaching their minima of 0.2129 and 0.1435, respectively.

As $\lambda$ increases, accuracy rises rapidly to around 64\%, yet the SSIM values also climb noticeably, indicating that the attacker recovers images structurally closer to the originals and privacy protection weakens. Notably, the SSIM curves exhibit a local dip around $\lambda=0.025$--$0.03$: at $\lambda=0.025$ the synthetic SSIM falls back to 0.2464, and at $\lambda=0.03$ the reconstruction SSIM and synthetic SSIM stay at relatively low levels (0.3278 and 0.2798), while accuracy already approaches its plateau at 64.31\%. Accuracy peaks at 65.53\% when $\lambda=0.035$ and then plateaus (64\%--65\%), but both SSIM metrics keep rising to about 0.44--0.45 for larger $\lambda$, meaning privacy protection deteriorates without further utility gain. Therefore, $\lambda=0.03$ achieves the optimal privacy-utility balance on CIFAR-10, confirming that $\lambda$ effectively steers the trade-off preference.

\subsection{Impact of Data Representation Ratio \texorpdfstring{$\rho$}{rho} on the Privacy--Utility Trade-off and System Cost}

This experiment answers one question --- how does the data representation ratio $\rho$ affect the privacy--utility trade-off and system cost?

\begin{wrapfigure}{l}{0.50\textwidth}
\centering
\includegraphics[width=\linewidth]{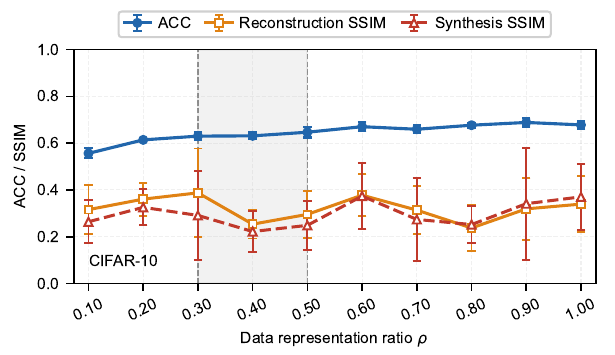}
\captionsetup{font=footnotesize,hypcap=false}
\caption{ACC improves near-linearly with $\rho$ while reconstruction leakage (SSIM) remains consistently low, confirming that capacity constraints effectively decouple utility preservation from information leakage across representation sizes.}
\label{fig:line_data_ratio_a}
\end{wrapfigure}
Experiments were conducted by scanning $\rho$ values from $0.10$ to $1.00$, where $\rho$ controls the proportion of data per category retained for training. The results (Figure~\ref{fig:line_data_ratio_a}) show that as $\rho$ increases, accuracy (ACC) improves from $55.63\%$ at $\rho=0.10$ to a peak of $68.85\%$ at $\rho=0.90$ ($67.81\%$ at $\rho=1.00$), with rapid gains up to $\rho=0.30$ ($63.00\%$) and slow, fluctuating growth thereafter.

In contrast, privacy protection, measured by the SSIM between the private images and the images recovered through reconstruction and synthesis attacks, remains consistently strong across all $\rho$ values, with reconstruction SSIM staying within $[0.24, 0.39]$ and synthesis SSIM within $[0.22, 0.37]$ and neither showing an upward trend, indicating that FedPix's directional information discarding mechanism works effectively regardless of data volume.

\newpage 
\begin{wrapfigure}{l}{0.50\textwidth}
\centering
\includegraphics[width=\linewidth]{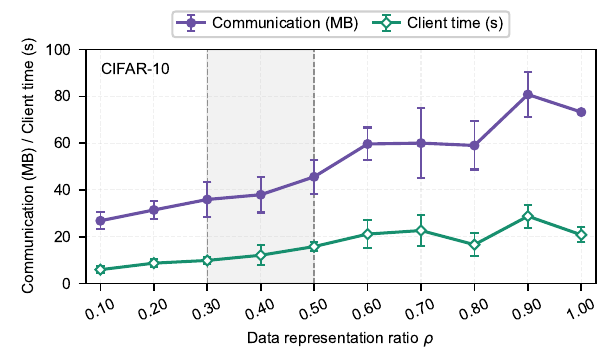}
\captionsetup{font=footnotesize,hypcap=false}
\caption{Total communication scales near-linearly with $\rho$, while client time remains low across all settings, confirming the lightweight client-side overhead of FedPix's data interface.}
\label{fig:line_data_ratio_b}
\end{wrapfigure}
Total communication volume scales nearly linearly with $\rho$ (Figure~\ref{fig:line_data_ratio_b}), from $26.81$~MB at $\rho=0.10$ to $80.80$~MB at $\rho=0.90$, while per-client training time increases from $5.90$ seconds to as much as $28.77$ seconds.

Considering all three dimensions, the optimal balance is achieved at $\rho \in [0.30, 0.50]$: for example, at $\rho=0.30$, ACC reaches $63.00\%$ ($93\%$ of the full-data accuracy), communication volume drops to $35.91$~MB ($\sim\!49\%$ of $\rho=1.00$), and client runtime reduces to $9.84$ seconds ($\sim\!47\%$ of $\rho=1.00$), validating that the effectiveness of capacity constraints does not depend on the amount of data.

\section{Encoder Convergence and System Scalability}
\label{sec:appendix_convergence_scalability}

\subsection{Encoder Training Convergence}

This experiment examines the convergence behavior of FedPix's server-side encoder training.

\begin{figure}[H]
\centering
\includegraphics[width=0.70\linewidth]{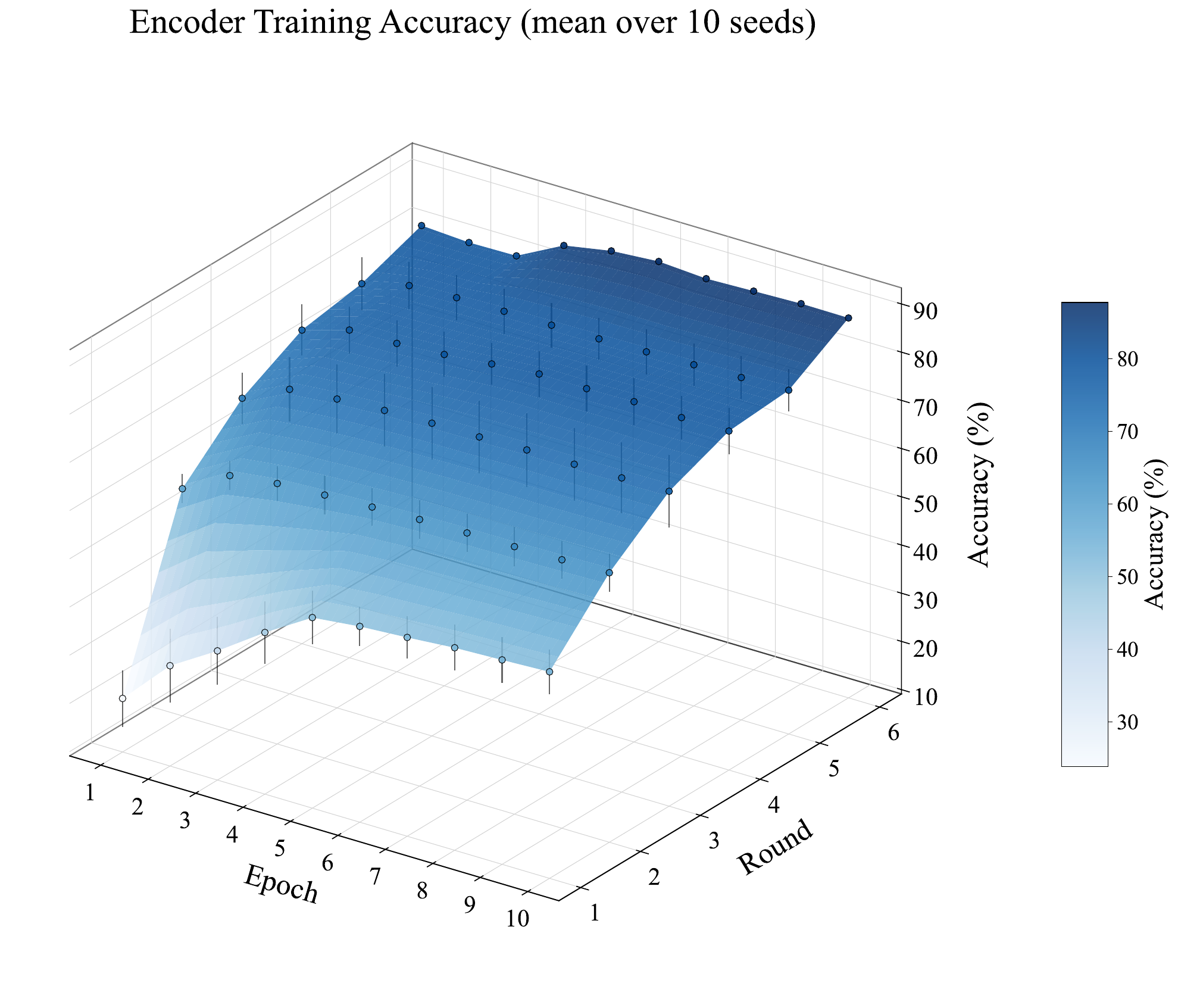}
\captionsetup{font=footnotesize,hypcap=false}
\caption{The encoder converges stably: within-round accuracy plateaus after 3--4 epochs, and the accuracy floor lifts from $\sim$22\% to $\sim$73\% over 6 rounds with variance narrowing to $\pm$1--2\%.}
\label{fig:encoder_convergence}
\end{figure}
During each communication round, the server trains the encoder $V_t$ for 10 epochs using the upsample--recompress closed-loop optimization described in Section~3. Figure~\ref{fig:encoder_convergence} visualizes the training accuracy on the consistent-sample set $\mathcal{I}_t^{\text{correct}}$ across 6 communication rounds and 10 training epochs per round, averaged over 10 random seeds on CIFAR-10.

Several observations emerge. First, within each round, accuracy rises rapidly during the first 3--4 epochs and then plateaus, indicating that the encoder quickly adapts to the current compressed representations and that further epochs yield diminishing returns. Second, across rounds, the accuracy floor consistently lifts: Round~1 starts at approximately 22\% and peaks near 51\%, while Round~6 starts above 73\% and reaches approximately 79\%. This upward shift reflects the progressive refinement of both the classifier $C_t$ and the compressor $V_t$ as more training data flows through the system.

Third, the standard deviation across seeds narrows substantially from earlier rounds (Round~1: $\pm$5--11\%) to later rounds (Round~6: $\pm$1--2\%), suggesting that the alternating optimization between classifier and encoder converges to a stable region regardless of initialization randomness. The smooth, monotonic surface with no visible regression confirms that the server-side closed-loop training provides stable and predictable convergence.

\subsection{System Scalability}

This experiment answers one question: \textbf{Can FedPix maintain stable performance as the number of clients scales up?} We evaluate on CIFAR-10 under four client counts (1, 5, 20, 100), comparing FedPix with four FL baselines (FedAvg, FedProx, SCAFFOLD, PARS) and two representation-learning based methods (FedSD2C, SFL).

\begin{table}[tbp]
\centering
\caption{FedPix accuracy remains nearly constant (65.8\%$\to$65.1\%) as client count grows from 1 to 100, while FL baselines degrade sharply (FedAvg: 78.5\%$\to$39.0\%), confirming that centralized training on pooled representations provides inherent scalability.}
\label{tab:scalability}
\begin{tabular}{lcccc}
\toprule[1.5pt]
\multirow{2}{*}{Method} & \multicolumn{4}{c}{ACC (\%) $\uparrow$} \\
\cmidrule(lr){2-5}
 & 1 Client & 5 Clients & 20 Clients & 100 Clients \\
\midrule
FedAvg (2017) & \textbf{78.52}$\pm$1.12 & 69.83$\pm$2.56 & 59.29$\pm$8.66 & 39.02$\pm$5.98 \\
FedProx (2020) & 76.68$\pm$0.90 & 68.87$\pm$2.93 & 61.82$\pm$7.14 & 32.84$\pm$3.48 \\
SCAFFOLD (2020) & 74.47$\pm$2.59 & \textbf{70.75}$\pm$2.40 & \textbf{68.14}$\pm$1.31 & 57.70$\pm$9.90 \\
PARS (2026) & 69.70$\pm$4.72 & 63.02$\pm$2.94 & 67.16$\pm$1.66 & 64.03$\pm$0.82 \\
FedSD2C (2024) & 28.72$\pm$1.36 & 39.69$\pm$2.96 & 45.54$\pm$0.72 & 44.22$\pm$0.84 \\
SFL (2020) & 72.01$\pm$1.84 & 44.79$\pm$5.77 & 57.41$\pm$3.55 & 51.27$\pm$8.24 \\
\rowcolor{gray!25} \textbf{FedPix (Ours)} & 65.79$\pm$0.97 & 65.84$\pm$1.23 & 65.44$\pm$1.26 & \textbf{65.14}$\pm$1.95 \\
\bottomrule[1.5pt]
\end{tabular}
\end{table}

Table~\ref{tab:scalability} reveals a clear scalability pattern. FL baselines (FedAvg, FedProx) uniformly exhibit steep accuracy declines as the client pool grows: FedAvg drops from 78.52\% (1 client) to 39.02\% (100 clients), and FedProx collapses from 76.68\% to 32.84\%, driven by the increasing difficulty of aggregating divergent local models under diminishing per-client data. SCAFFOLD partially mitigates this drift via control variates, retaining 57.70\% at 100 clients, while PARS remains notably resilient (64.03\% at 100 clients) owing to its adaptive pseudo-labeling strategy. SFL shows strong single-client performance (72.01\%), benefiting from the full dataset being available to one split-learning instance, but degrades to 51.27\% at 100 clients as feature-space alignment across many split paths becomes challenging. FedSD2C remains consistently low across all scales (28.72--45.54\%), reflecting the limited utility of its latent codes under data fragmentation.

In stark contrast, \textbf{FedPix maintains nearly constant accuracy} across all client scales: 65.79\% (1 client) to 65.14\% (100 clients), a drop of only 0.65 percentage points with uniformly tight standard deviations ($\pm$0.97--1.95\%). At 20 and 100 clients, FedPix outperforms the strongest baseline by 0.6 and 1.1 percentage points, respectively, while at 100 clients it surpasses FedAvg by over 26 percentage points. This stability arises from the same architectural property that confers Non-IID robustness (see Non-IID experiment in the main text): the server trains a single centralized classifier on pooled compressed representations, so adding more clients simply increases the diversity and volume of the training pool without aggravating the challenge of aggregating divergent models. The server’s classifier operates on a unified, fixed-dimensional representation space, effectively decoupling the final performance from the number of participating clients. The unified compressor design thus provides \textbf{inherent scalability}: performance depends on the total data available to the server, not on how that data is partitioned across clients.

\section{Non-IID Sensitivity Analysis}
\label{sec:appendix_noniid}

\subsection{Robustness to Non-IID Data Distributions}

We evaluate robustness to heterogeneous client distributions under four Dirichlet concentration parameters ($\alpha \in \{0.01, 0.1, 0.3, 1.0\}$)~\citep{zhao2018federated,li2020federated}.

\begin{table}[tbp]
\centering
\caption{FedPix maintains stable accuracy (63--65\%) across all $\alpha$, while traditional FL baselines collapse to near-random accuracy under extreme heterogeneity ($\alpha=0.01$), confirming that data-level transmission inherently decouples client data distribution from server-side training.}
\label{tab:non_iid_performance}
\footnotesize
\begin{tabular}{l|c|c|c|c}
\toprule[1.5pt]
\multirow{2}{*}{Method} & \multicolumn{4}{c}{ACC (\%) $\uparrow$} \\
\cmidrule(lr){2-5}
 & $\alpha=0.01$ & $\alpha=0.1$ & $\alpha=0.3$ & $\alpha=1.0$ \\
\midrule
FedAvg (2017) & 11.09$\pm$1.74 & 22.87$\pm$9.58 & 60.74$\pm$9.57 & 71.28$\pm$0.61 \\
FedProx (2020) & 11.39$\pm$2.46 & 24.02$\pm$10.22 & 60.89$\pm$8.20 & 71.23$\pm$0.73 \\
SCAFFOLD (2020) & 10.22$\pm$0.35 & 18.93$\pm$19.62 & 54.02$\pm$22.85 & \textbf{73.28$\pm$0.77} \\
PARS (2026) & 9.94$\pm$0.20 & 21.49$\pm$14.01 & 48.62$\pm$23.37 & 58.15$\pm$26.99 \\
FedSD2C (2024) & 29.88$\pm$1.86 & 36.69$\pm$1.12 & 41.95$\pm$2.12 & 47.81$\pm$1.40 \\
SFL (2020) & 11.52$\pm$3.32 & 26.58$\pm$7.89 & 43.29$\pm$8.91 & 53.22$\pm$8.52 \\
\rowcolor{gray!25} \textbf{FedPix (Ours)} & \textbf{64.72$\pm$1.79} & \textbf{63.42$\pm$4.18} & \textbf{64.65$\pm$0.48} & 65.34$\pm$1.00 \\
\bottomrule[1.5pt]
\end{tabular}
\end{table}

Traditional FL methods (FedAvg, FedProx, SCAFFOLD, PARS) collapse under extreme heterogeneity---at $\alpha=0.01$, all fall to 9.9--11.4\%, near random guessing for CIFAR-10 (10\%). FedPix maintains stable accuracy across the entire spectrum (63.42--65.34\%), with only a 0.62\% drop from $\alpha=1.0$ to $\alpha=0.01$.

This confirms that the data-interface paradigm inherently decouples client heterogeneity from server-side training: the server trains on pooled representations rather than aggregating locally-trained models, and the learned encoder preserves task-relevant structure even under severe distribution shift.

The experimental results above demonstrate a consistent pattern---capacity-constrained representations achieve strong reconstruction resistance while preserving classification accuracy. The following sections provide a formal theoretical framework that explains this phenomenon, along with complete proofs.

\section{Theoretical Analysis}
\label{sec:theory}

\subsection{Problem Setup}
\label{sec:problem-setup}

Consider $K$ clients, where client $k$ holds a local dataset
$
\mathcal D_k
=
\{(D_{k,i},L_{k,i})\}_{i=1}^{n_k},
$
with
$
D_{k,i}\in[0,1]^{S\times S\times C}
$
and
$
L_{k,i}\in\{1,\ldots,Q\}.
$
Let
$
N=\sum_{k=1}^{K}n_k
$
and
$
p_k=n_k/N.
$

FedPix uses a shared representation encoder
$
V:
[0,1]^{S\times S\times C}
\rightarrow
[0,1]^{y\times y\times C},
\qquad y<S,
$
where the output size $y$ controls the physical capacity of the
uploaded representation. For an input $D_{k,i}$, the client uploads
$
I_{k,i}=V(D_{k,i})
$
together with its label $L_{k,i}$. Before classification, the server
uses the fixed structure-preserving upsampling operator
$\operatorname{UpSample}(\cdot)$ to restore $I_{k,i}$ to spatial size
$S\times S$. Let $C$ denote the server-side classifier.

The utility objective is the classification loss on the uploaded
representations:
\begin{equation}
\mathcal L_{\mathrm{cls}}(C,V)
=
\sum_{k=1}^{K}
p_k
\frac{1}{n_k}
\sum_{i=1}^{n_k}
\ell_{\mathrm{ce}}
\left(
C\!\left(
\operatorname{UpSample}(V(D_{k,i}))
\right),
L_{k,i}
\right).
\label{eq:global-classification-loss}
\end{equation}

To suppress visually reconstructable information, we measure the
similarity between an original image and its uploaded representation
after upsampling:
\begin{align}
\mathcal L_{\mathrm{sim}}(V)
=
\sum_{k=1}^{K}
p_k
\frac{1}{n_k}
\sum_{i=1}^{n_k}
\Big[
&
\operatorname{SSIM}
\left(
\operatorname{UpSample}(V(D_{k,i})),
D_{k,i}
\right)
\nonumber\\
&+
\beta\,
\operatorname{PSNR}
\left(
\operatorname{UpSample}(V(D_{k,i})),
D_{k,i}
\right)
\Big].
\label{eq:global-similarity-loss}
\end{align}
Lower $\mathcal L_{\mathrm{sim}}(V)$ indicates lower direct pixel-level and structural similarity to the original images; it does not by itself bound information leakage to an adaptive attacker. This raw-data objective is an analytical formulation. The server instead optimizes the recompression surrogate in Section~\ref{subsec:full_iteration}, and the analysis does not establish equivalence between the two objectives.

FedPix seeks a compressor that minimizes reconstruction similarity
while maintaining classification utility. We formulate this goal as
\begin{equation}
\begin{aligned}
\min_{C,V}\quad
& \mathcal L_{\mathrm{sim}}(V) \\
\text{s.t.}\quad
& \mathcal L_{\mathrm{cls}}(C,V)
\leq \epsilon,
\end{aligned}
\label{eq:privacy-utility-problem}
\end{equation}
where $\epsilon$ specifies the maximum admissible classification loss.
A Lagrangian relaxation of
Problem~\eqref{eq:privacy-utility-problem} is
$
\min_{C,V}
\quad
\lambda\,
\mathcal L_{\mathrm{cls}}(C,V)
+
\mathcal L_{\mathrm{sim}}(V),
$
where $\lambda>0$ controls the privacy--utility trade-off.

The optimization strategy for solving Problem~\eqref{eq:privacy-utility-problem} under federated constraints is described in the main text (Section~\ref{subsec:full_iteration}). We now provide the formal theoretical analysis.

In this section, $C_{t+1}$ denotes the pre-softmax logit map of the
classifier trained on the server. For a logit vector $z$ and label
$l$, let
$
\ell_{\mathrm{cls}}(z,l)
=
-\log[\operatorname{softmax}(z)]_l.
$
For the original image $D_t$ and its upsampled compressed
representation $I_t^{\mathrm{up}}$, define
$
\Delta_t=I_t^{\mathrm{up}}-D_t.
$

\begin{assumption}[Classifier smoothness]
\label{ass:local-classifier-smoothness}
There exists $\kappa>0$ such that, for any $u,v$ on the line segment
between $D_t$ and $I_t^{\mathrm{up}}$,
$
\left\|
\nabla_D C_{t+1}(u)-\nabla_D C_{t+1}(v)
\right\|_{\mathrm{op}}
\leq
\kappa\|u-v\|_2,
$
where $\nabla_D C_{t+1}$ is the Jacobian of the classifier logits with
respect to the input image.
\end{assumption}

\begin{proposition}[Classifier-sensitive distortion]
\label{prop:sensitive-distortion}
Under Assumption~\ref{ass:local-classifier-smoothness}, for a sample
$(D_t,L_t)$,
\begin{align}
&\left|
\ell_{\mathrm{cls}}
\left(C_{t+1}(I_t^{\mathrm{up}}),L_t\right)
-
\ell_{\mathrm{cls}}
\left(C_{t+1}(D_t),L_t\right)
\right|
\nonumber\\
&\quad\leq
\sqrt{2}\left(
\left\|
\nabla_D C_{t+1}(D_t)\Delta_t
\right\|_2
+
\frac{\kappa}{2}\|\Delta_t\|_2^2
\right).
\label{eq:sensitive-distortion-bound}
\end{align}
\end{proposition}

Proposition~\ref{prop:sensitive-distortion} separates pixel distortion
from classifier-sensitive distortion. Thus, a compressed image may
have substantially lower PSNR and SSIM than the original image while
incurring only a small utility loss, provided that its distortion has
limited effect on the classifier logits. This establishes the
possibility of visually distorted yet task-useful representations.

We next isolate the role of upload size using a canonical linear model. This idealized Gaussian model is separate from the bounded pixel domain above and is not a guarantee for the nonlinear encoder or its closed-loop training dynamics.
After vectorization, let
$
x_t=\operatorname{vec}(D_t)\in\mathbb R^d,
d=CS^2,
$
and let
$
r_y=Cy^2
$
denote the dimension of an uploaded image of size $C\times y\times y$.
For any representation $Z$, $H(L_t\mid Z)$ denotes the minimum expected
cross-entropy attainable by a probabilistic classifier using $Z$:
\begin{equation}
H(L_t\mid Z)
=
\inf_q
\mathbb E[-\log q(L_t\mid Z)].
\label{eq:conditional-entropy-ce}
\end{equation}

\begin{assumption}[Canonical task--nuisance model]
\label{ass:canonical-task-subspace}
The vectorized image admits the orthogonal decomposition
\begin{equation}
x_t
=
U_{\mathrm{task}}a_t
+
U_{\mathrm{nui}}b_t,
\label{eq:task-nuisance-decomposition}
\end{equation}
where
$U_{\mathrm{task}}\in\mathbb R^{d\times d_{\mathrm{task}}}$
and
$U_{\mathrm{nui}}\in
\mathbb R^{d\times(d-d_{\mathrm{task}})}$
have orthonormal and mutually orthogonal columns. The task component
$a_t$ is a minimal linear representation preserving all label-relevant
information:
$
p(L_t\mid x_t)=p(L_t\mid a_t),
$
and no linear representation of dimension smaller than
$d_{\mathrm{task}}$ has this property. Moreover,
$
a_t\sim\mathcal N(0,I_{d_{\mathrm{task}}}),
\qquad
b_t\sim\mathcal N(0,I_{d-d_{\mathrm{task}}}),
$
with $b_t$ independent of $(a_t,L_t)$.
\end{assumption}

The task component contains information needed for classification,
whereas the nuisance component is label-independent but may still aid
image reconstruction. We model compression and upsampling by a
rank-$r_y$ orthogonal projection $P_y$ and define the normalized Bayes
reconstruction error
\begin{equation}
\mathcal D_{\mathrm{rec}}(P_y)
=
\frac{1}{d}
\mathbb E
\left[
\left\|
x_t-\mathbb E[x_t\mid P_yx_t,L_t]
\right\|_2^2
\right].
\label{eq:reconstruction-distortion}
\end{equation}
The estimator conditions on $L_t$ because the label is uploaded together
with the compressed representation.

\begin{theorem}[Task-preserving capacity threshold]
\label{thm:capacity-threshold}
Under Assumption~\ref{ass:canonical-task-subspace}, if
$r_y\geq d_{\mathrm{task}}$, there exists a rank-$r_y$
orthogonal projection $P_y$ that preserves the complete task subspace
and satisfies
$
H(L_t\mid P_yx_t)=H(L_t\mid x_t).
$
If its remaining $r_y-d_{\mathrm{task}}$ dimensions retain arbitrary
nuisance directions, then
$
\mathcal D_{\mathrm{rec}}(P_y)
=
1-\frac{r_y}{d}.
$
When $r_y<d_{\mathrm{task}}$, no linear representation of dimension
$r_y$ can preserve all label-relevant information.
\end{theorem}

Theorem~\ref{thm:capacity-threshold} identifies a task-preserving
representation capacity region. As long as $r_y\geq d_{\mathrm{task}}$,
nuisance directions can be discarded, increasing reconstruction error
without increasing the Bayes-optimal classification loss. Below this
threshold, some task-relevant information must also be removed.

\section{Proofs for the Theoretical Analysis}
\label{app:theory-proofs}

\subsection{Proof of \texorpdfstring{Proposition~\ref{prop:sensitive-distortion}}{the Classifier-Sensitive Distortion Proposition}}

For a logit vector $z\in\mathbb R^Q$ and label $l$, the softmax
cross-entropy is
\begin{equation}
\ell_{\mathrm{cls}}(z,l)
=
-z_l+\log\sum_{c=1}^{Q}\exp(z_c).
\end{equation}
Its gradient with respect to the logits is
\begin{equation}
\nabla_z\ell_{\mathrm{cls}}(z,l)
=
\operatorname{softmax}(z)-e_l,
\end{equation}
where $e_l$ is the one-hot vector for class $l$. Since both
$\operatorname{softmax}(z)$ and $e_l$ are probability vectors,
\begin{equation}
\left\|\nabla_z\ell_{\mathrm{cls}}(z,l)\right\|_2
\leq\sqrt{2}.
\end{equation}
The mean-value inequality therefore gives, for arbitrary logits
$z,z'$,
\begin{equation}
\left|
\ell_{\mathrm{cls}}(z',l)-\ell_{\mathrm{cls}}(z,l)
\right|
\leq
\sqrt{2}\|z'-z\|_2.
\label{eq:ce-logit-lipschitz}
\end{equation}

Recall that
$I_t^{\mathrm{up}}=D_t+\Delta_t$. By the fundamental theorem of
calculus,
\begin{align}
&C_{t+1}(I_t^{\mathrm{up}})-C_{t+1}(D_t)
\nonumber\\
&\quad=
\int_0^1
\nabla_D C_{t+1}(D_t+s\Delta_t)\Delta_t\,ds
\nonumber\\
&\quad=
\nabla_D C_{t+1}(D_t)\Delta_t+R_t,
\label{eq:logit-taylor}
\end{align}
where
\begin{equation}
R_t
=
\int_0^1
\left[
\nabla_D C_{t+1}(D_t+s\Delta_t)
-\nabla_D C_{t+1}(D_t)
\right]
\Delta_t\,ds.
\end{equation}
Assumption~\ref{ass:local-classifier-smoothness} implies
\begin{align}
\|R_t\|_2
&\leq
\int_0^1
\kappa s\|\Delta_t\|_2^2\,ds
\nonumber\\
&=
\frac{\kappa}{2}\|\Delta_t\|_2^2.
\label{eq:taylor-remainder-bound}
\end{align}
Applying Eq.~\eqref{eq:ce-logit-lipschitz} to the two classifier
outputs and then using Eqs.~\eqref{eq:logit-taylor} and
\eqref{eq:taylor-remainder-bound} yields
\begin{align}
&\left|
\ell_{\mathrm{cls}}
\left(C_{t+1}(I_t^{\mathrm{up}}),L_t\right)
-
\ell_{\mathrm{cls}}
\left(C_{t+1}(D_t),L_t\right)
\right|
\nonumber\\
&\quad\leq
\sqrt{2}\left(
\left\|
\nabla_D C_{t+1}(D_t)\Delta_t
\right\|_2
+
\frac{\kappa}{2}\|\Delta_t\|_2^2
\right),
\end{align}
which proves Proposition~\ref{prop:sensitive-distortion}.
\hfill$\square$

\subsection{Proof of \texorpdfstring{Theorem~\ref{thm:capacity-threshold}}{the Task-Preserving Capacity Theorem}}

Suppose first that $r_y\geq d_{\mathrm{task}}$, and let
$k=r_y-d_{\mathrm{task}}$. Choose any matrix
$U_{\mathrm{keep}}\in\mathbb R^{d\times k}$ whose columns are
orthonormal nuisance directions in
$\operatorname{span}(U_{\mathrm{nui}})$, and define
\begin{equation}
P_y
=
U_{\mathrm{task}}U_{\mathrm{task}}^\top
+
U_{\mathrm{keep}}U_{\mathrm{keep}}^\top.
\label{eq:task-preserving-projection}
\end{equation}
Then $P_y$ is an orthogonal projection of rank
$d_{\mathrm{task}}+k=r_y$. Moreover,
\begin{equation}
U_{\mathrm{task}}^\top P_yx_t
=
U_{\mathrm{task}}^\top x_t
=a_t,
\end{equation}
so the complete task component can be recovered from $P_yx_t$.
Because $P_yx_t$ is a function of $x_t$, while $a_t$ is a function of
$P_yx_t$, conditioning reduces entropy in both directions:
\begin{equation}
H(L_t\mid x_t)
\leq
H(L_t\mid P_yx_t)
\leq
H(L_t\mid a_t).
\end{equation}
Assumption~\ref{ass:canonical-task-subspace} gives
$H(L_t\mid a_t)=H(L_t\mid x_t)$, and hence
\begin{equation}
H(L_t\mid P_yx_t)=H(L_t\mid x_t).
\label{eq:utility-preservation}
\end{equation}

To compute the reconstruction error, complete
$U_{\mathrm{keep}}$ with a matrix $U_{\mathrm{drop}}$ so that
$[U_{\mathrm{keep}},U_{\mathrm{drop}}]$ is an orthonormal basis of the
nuisance subspace. The decomposition in
Assumption~\ref{ass:canonical-task-subspace} can then be written as
\begin{equation}
x_t
=
P_yx_t+U_{\mathrm{drop}}b_t^{\mathrm{drop}},
\end{equation}
where $b_t^{\mathrm{drop}}$ is a standard Gaussian vector of dimension
$d-r_y$. By the independence and isotropy assumptions,
$U_{\mathrm{drop}}b_t^{\mathrm{drop}}$ is zero mean and independent of
$(P_yx_t,L_t)$. Consequently,
\begin{equation}
\mathbb E[x_t\mid P_yx_t,L_t]=P_yx_t.
\end{equation}
Using the orthonormality of $U_{\mathrm{drop}}$ gives
\begin{align}
\mathcal D_{\mathrm{rec}}(P_y)
&=
\frac{1}{d}
\mathbb E
\left\|
U_{\mathrm{drop}}b_t^{\mathrm{drop}}
\right\|_2^2
\nonumber\\
&=
\frac{d-r_y}{d}
=
1-\frac{r_y}{d}.
\label{eq:reconstruction-capacity-proof}
\end{align}

Now suppose that $r_y<d_{\mathrm{task}}$ and consider any linear
representation $Z=Bx_t$ with $\operatorname{rank}(B)\leq r_y$. Since
$Z$ is a deterministic function of $x_t$,
\begin{equation}
H(L_t\mid Z)-H(L_t\mid x_t)
=
I(L_t;x_t\mid Z)
\geq0.
\end{equation}
Equality would mean that $Z$ preserves all label-relevant information
in $x_t$. However, its rank is strictly smaller than
$d_{\mathrm{task}}$, contradicting the minimality condition in
Assumption~\ref{ass:canonical-task-subspace}. Therefore,
\begin{equation}
H(L_t\mid Bx_t)>H(L_t\mid x_t).
\end{equation}
By the conditional-entropy characterization in
Eq.~\eqref{eq:conditional-entropy-ce}, every such representation has
strictly larger Bayes-optimal cross-entropy. This completes the proof.
\hfill$\square$

\section{On-Device Encoder Deployment and Benchmark}
\label{sec:appendix_ondevice}

To validate that the FunnelEncoder can operate on resource-limited client devices, we deployed the encoder on a commercial Android device with Snapdragon~660 SoC (2018) and benchmarked it with 100 real-world photographs. The encoder was exported to ONNX Runtime Mobile (arm64-v8a, FP32). Each photograph ($\sim$12\,MP) is resized to $128 \times 128$ and normalized to $[-1, 1]$ before inference, producing a $3 \times 8 \times 8$ output (768 bytes). Table~\ref{tab:encoder_config} summarizes the encoder configuration; Table~\ref{tab:device_specs} lists the benchmark device specifications.

\begin{table}[H]
\centering
\caption{Encoder configuration for on-device benchmark.}
\label{tab:encoder_config}
\begin{tabular}{lrl}
\toprule[1.5pt]
\textbf{Parameter} & \textbf{Value} & \textbf{Description} \\
\midrule
Input channels   & 3    & RGB input \\
Resize target     & $128 \times 128$ & Pre-resize from native photo \\
Encoder input     & $128 \times 128$ & Fixed input resolution \\
Output size       & $8 \times 8$    & Compressed spatial size \\
Hidden dimension  & 128  & Internal channel count \\
Convs per block   & 5    & ConvPoolBlock depth \\
Compress to RGB   & True & 3-channel output (interpretable) \\
Output format     & $3 \times 8 \times 8$ & 768 bytes (FP32) \\
Total parameters  & $\sim$1.35M & Model $\approx$ 2.3 MB (ONNX) \\
\bottomrule[1.5pt]
\end{tabular}
\end{table}
The encoder uses a fixed \texttt{AveragePool} ($\text{kernel}=\text{stride}=16$) in place of \texttt{AdaptiveAvgPool2d} to eliminate dynamic shape resolution overhead on mobile. The $128 \times 128$ input balances visual fidelity, $\sim$1.35\,M parameter footprint ($<$1\,s per image on Snapdragon~660), and RAM constraints ($\sim$2\,GB available). Larger inputs quadratically increase FLOPs and memory; smaller inputs discard too much spatial detail pre-compression.

\begin{table}[H]
\centering
\caption{Hardware specifications of the benchmark device.}
\label{tab:device_specs}
\begin{tabular}{ll}
\toprule[1.5pt]
\textbf{Attribute} & \textbf{Value} \\
\midrule
Manufacturer / Model & Commercial device (Snapdragon~660) \\
SoC / Board          & Snapdragon 660 (sdm660) \\
CPU architecture     & arm64-v8a, 8 cores @ 2.21 GHz \\
Total / Available RAM & 5.70 GB / 2.02 GB \\
Total / Available Storage & 110.4 GB / 6.77 GB \\
GPU Renderer         & Adreno 512 \\
Android version      & 9 (SDK 28) \\
Screen               & $1080 \times 2201$ px \\
\bottomrule[1.5pt]
\end{tabular}
\end{table}
The benchmark device (2018) provides a concrete example of CPU-only deployment on older mobile hardware. The measured results apply to this device and workload; performance on other devices, including NPU-equipped phones, requires separate benchmarking.

\subsection{Benchmark Results}

\begin{table}[H]
\renewcommand{\arraystretch}{0.95}
\centering
\caption{End-to-end on-device encoder benchmark results (100 photographs).}
\label{tab:benchmark_results}
\begin{tabular}{lr}
\toprule[1.2pt]
\textbf{Metric} & \textbf{Value} \\
\midrule
Total images processed   & 100 \\
Total wall-clock time    & 85.62 s \\
Avg.\ encode time / img  & 849.33 ms \\
Throughput (images/sec)  & 1.17 img/s \\
Throughput (input data)  & 3.83 MB/s \\
Total input data         & 328.22 MB \\
Total compressed output  & 0.0768 MB (76.8 KB) \\
Representation reduction  & 4,274$\times$ \\
Per-image output         & 768 bytes \\
\bottomrule[1.2pt]
\end{tabular}
\end{table}
All 100 images completed without a reported runtime failure. Average encode time is 849\,ms (dominated by the convolutional forward pass). Throughput of 1.17\,img/s means 1,000 images compress in $\sim$14.2 minutes. The pipeline reduces transmitted representation size by approximately $4,274\times$ (328.22\,MB $\to$ 0.0768\,MB, using decimal units), reducing per-image communication to a fixed 768 bytes.

\begin{table}[H]
\renewcommand{\arraystretch}{0.95}
\centering
\caption{Battery and thermal measurements before and after the benchmark.}
\label{tab:battery_analysis}
\begin{tabular}{lccc}
\toprule[1.2pt]
\textbf{Metric} & \textbf{Before} & \textbf{After} & \textbf{Delta} \\
\midrule
Battery level          & 97.0\% & 94.0\% & $-3.0$ pp \\
Battery voltage        & 4,089 mV & 3,849 mV & $-240$ mV \\
Charge counter         & 1,859 $\mu$Ah & 1,832 $\mu$Ah & $-27$ $\mu$Ah \\
Device temperature     & 31.5$^\circ$C & 32.0$^\circ$C & $+0.5^\circ$C \\
Charging state         & \multicolumn{3}{c}{Not charging (battery-only)} \\
\bottomrule[1.2pt]
\end{tabular}
\end{table}
The benchmark was conducted on battery power. The reported battery level changed from 97.0\% to 94.0\%, and device temperature increased by $0.5^\circ$C. The charge-counter readings and their units require validation before energy per image can be estimated; these endpoint readings alone do not establish the absence of thermal throttling.

\medskip
\noindent\textbf{Discussion.} The encoder runs on the Snapdragon~660 device with CPU-only FP32 inference at 1.17\,images/s. The client performs a forward pass per sample while encoder training remains on the server. Each transmitted representation contains 192 FP32 values (768 bytes), a $256\times$ reduction relative to a $3\times128\times128$ FP32 input tensor. The approximately $4,274\times$ ratio above instead compares the total source photograph files with representation payloads; it excludes encoder downloads, labels, and protocol overhead. INT8 quantization, dynamic input resolution, and NPU acceleration remain future work, with their effects on utility, latency, and energy still to be measured.

\section{Adaptive Reconstruction Attack}
\label{sec:appendix_adaptive_attack}

To evaluate the reconstruction resistance of FedPix's capacity bottleneck under a strong adversary, we design a \textbf{white-box adaptive reconstruction attack}. The attacker trains a dedicated restoration network to invert the compression mapping, with full knowledge of the encoder architecture and parameters.

\subsection{Attack Setup}

We consider an attacker with the following capabilities:
\begin{enumerate}
    \item \textbf{Access to compressed representations.} The attacker observes the transmitted compressed images $\mathbf{X} = E(\mathbf{y})$, where $E$ is the FunnelEncoder and $\mathbf{y}$ is the original image.
    \item \textbf{Full encoder knowledge (white-box).} The attacker possesses the complete encoder architecture and its trained parameters. This models the scenario where the encoder is standardized and publicly released, or where parameters are obtained through system-level access.
    \item \textbf{Shadow dataset.} The attacker holds an auxiliary dataset $\mathcal{D}_{\text{shadow}}$ drawn from the same data distribution, used exclusively for training the restoration model.
\end{enumerate}

The attacker trains a restoration model $R_\phi: \mathcal{X} \to \mathcal{Y}$ to minimize the $\ell_1$ reconstruction loss over the shadow dataset:
\begin{equation}
    \min_\phi \; \mathbb{E}_{(\mathbf{x}, \mathbf{y}) \sim \mathcal{D}_{\text{shadow}}} \left[ \| R_\phi(E(\mathbf{y})) - \mathbf{y} \|_1 \right],
\end{equation}
where $E$ is frozen with the known parameters. After training, $R_\phi$ is applied to the actual transmitted set $\mathbf{X}$, producing reconstructed estimates $\hat{\mathbf{Y}} = R_\phi(\mathbf{X})$, which are compared against the ground-truth originals $\mathbf{Y}$.

\subsection{Restoration Model}

The restoration network is a \textbf{ResUNet}---a U-Net augmented with residual bottleneck blocks. The encoder path contains four down-sampling stages (DoubleConv $\to$ MaxPool, base channels 64), expanding to 512 channels at $8 \times 8$ spatial resolution. The bottleneck consists of two residual blocks (Conv$\to$BN$\to$ReLU$\to$Conv$\to$BN with skip connection). The decoder path mirrors the encoder with four up-sampling stages and skip connections, followed by a $1\times 1$ convolution with $\tanh$ activation. The model has $\sim$500K trainable parameters. Training uses Adam ($\text{lr}=10^{-3}$) with cosine annealing over 20 epochs, minimizing $\ell_1$ loss.

\subsection{Results}

We evaluate on the same four datasets and use the same PSNR / SSIM / LPIPS metrics defined in Section~\ref{sec:appendix_metrics}. Table~\ref{tab:adaptive_attack_results} reports the reconstruction quality of the ResUNet attacker across all four datasets, alongside FedPix's Synthetic and Reconstructed metrics from the main baseline comparison in Section~\ref{sec:experiments}. All three rows per dataset share the same compressed representations; they differ only in whether an active restoration step is applied and how strong that restoration model is.

\begin{table}[tbp]
\centering
\caption{Adaptive reconstruction attack results under the white-box setting. Best value in each column is \textbf{bolded}. $\downarrow$ = lower is better for reconstruction resistance; $\uparrow$ = higher is better for reconstruction resistance.}
\label{tab:adaptive_attack_results}
\begin{tabular}{llccc}
\toprule[1.5pt]
\textbf{Dataset} & \textbf{Attack Type} & \textbf{PSNR (dB) $\downarrow$} & \textbf{SSIM $\downarrow$} & \textbf{LPIPS $\uparrow$} \\
\midrule
\multirow{3}{*}{BloodMNIST}
    & Adaptive (ResUNet)   & 13.53 $\pm$ 1.07 & 0.48 $\pm$ 0.07 & 0.11 $\pm$ 0.02 \\
    & Synthetic            & \textbf{8.84 $\pm$ 1.34}  & \textbf{0.32 $\pm$ 0.10} & \textbf{0.26 $\pm$ 0.02} \\
    & Reconstructed (MoIA) & 10.20 $\pm$ 1.02 & 0.34 $\pm$ 0.08 & \textbf{0.26 $\pm$ 0.03} \\
\midrule
\multirow{3}{*}{FashionMNIST}
    & Adaptive (ResUNet)   & 13.81 $\pm$ 1.88 & 0.56 $\pm$ 0.06 & 0.08 $\pm$ 0.02 \\
    & Synthetic            & 8.27 $\pm$ 2.34  & 0.29 $\pm$ 0.09 & 0.39 $\pm$ 0.09 \\
    & Reconstructed (MoIA) & \textbf{8.19 $\pm$ 0.98}  & \textbf{0.26 $\pm$ 0.05} & \textbf{0.41 $\pm$ 0.05} \\
\midrule
\multirow{3}{*}{CIFAR-10}
    & Adaptive (ResUNet)   & 11.34 $\pm$ 1.49 & 0.44 $\pm$ 0.12 & 0.19 $\pm$ 0.05 \\
    & Synthetic            & \textbf{8.47 $\pm$ 1.44}  & \textbf{0.24 $\pm$ 0.07} & \textbf{0.39 $\pm$ 0.05} \\
    & Reconstructed (MoIA) & 10.94 $\pm$ 1.12 & 0.33 $\pm$ 0.08 & 0.30 $\pm$ 0.08 \\
\midrule
\multirow{3}{*}{MNIST}
    & Adaptive (ResUNet)   & 10.80 $\pm$ 2.59 & 0.38 $\pm$ 0.12 & 0.09 $\pm$ 0.03 \\
    & Synthetic            & \textbf{6.74 $\pm$ 2.32}  & 0.19 $\pm$ 0.08 & 0.42 $\pm$ 0.08 \\
    & Reconstructed (MoIA) & 6.87 $\pm$ 1.46  & \textbf{0.11 $\pm$ 0.03} & \textbf{0.43 $\pm$ 0.07} \\
\bottomrule[1.5pt]
\end{tabular}
\end{table}
\textbf{Active restoration recovers additional information.} The Adaptive (ResUNet) attacker achieves PSNR of 10.80--13.81\,dB, improving on the direct Synthetic comparison by 2.87--5.54\,dB. Relative to MoIA, its PSNR gains range from 0.40\,dB on CIFAR-10 to 5.62\,dB on FashionMNIST, accompanied by higher SSIM and lower LPIPS on every dataset. Thus, direct representation dissimilarity can understate recoverable information. These metrics do not define a universal threshold for usable reconstruction.

\textbf{Global structure partially survives compression.} LPIPS values for the Adaptive attacker are moderate (0.08--0.19), indicating that the bottleneck retains coarse spatial layout and color statistics---precisely the information needed for downstream classification. The tension between low PSNR and moderate LPIPS reflects the design intent of FedPix: preserving \textit{what} is in the image while discarding \textit{how it looks} in detail.

\textbf{Scope of the attack evaluation.} These results characterize the specified white-box restoration model and training budget. They are not an upper bound on attacker success; other architectures, training budgets, or auxiliary data could yield stronger reconstructions.

\subsection{Analysis}

\textbf{Dataset-dependent reconstruction quality.} CIFAR-10 has the highest Adaptive LPIPS (0.19), whereas MNIST has the lowest Adaptive PSNR (10.80\,dB, versus 11.34\,dB on CIFAR-10). Different metrics therefore yield different dataset rankings; these results do not establish that natural images are uniformly the most resistant to reconstruction.

\textbf{Input complexity has limited impact.} Across four datasets spanning simple digits (MNIST) to natural scenes (CIFAR-10), the Adaptive PSNR range is narrow (10.80--13.81\,dB). The $8 \times 8$ bottleneck dominates over input complexity: even MNIST digits on blank backgrounds remain substantially degraded after reconstruction. This suggests that reconstruction resistance is a structural property of the spatial reduction ratio rather than dataset-dependent.

\textbf{White-box attack remains relevant.} Across all four datasets, the learned attacker improves PSNR, SSIM, and LPIPS relative to MoIA. These gains show that a spatial bottleneck reduces representation size without necessarily eliminating exploitable structure. Reconstruction resistance must therefore be assessed against adaptive attacks as well as direct image similarity.

\section{Related Work}\label{sec:rela}

\subsection{Traditional Federated Learning}

Federated learning enables collaborative training without sharing raw data and has become a foundational paradigm for privacy-sensitive services~\citep{mcmahan2017communication}. FedAvg~\citep{mcmahan2017communication}, as the earliest and most widely adopted aggregation algorithm, constructs a global model by weighted averaging of client updates but suffers from severe client drift under non-IID data distributions. FedProx~\citep{li2020fedprox} introduces a proximal term to constrain local updates, SCAFFOLD~\citep{karimireddy2020scaffold} corrects update biases via control variates, FedAdam~\citep{reddi2020adaptive} brings adaptive optimization to federated learning, and PARS~\citep{pars2026} designs pseudo-labeling with adaptive sampling strategies for semi-supervised scenarios. FedLADA~\citep{sun2023efficient} incorporates a locally amended technique into adaptive optimizers to mitigate client drift, while $\Delta$-FL~\citep{pillutla2024federated} employs superquantile aggregation to improve performance across heterogeneous clients.

A common thread across these methods is the \textit{model parameter exchange} paradigm: each client locally optimizes on its own data distribution $\mathcal{D}_k$, yielding a local model $w_k$. When $\mathcal{D}_k \neq \mathcal{D}_j$, local objectives diverge, causing client drift and aggregation bias. Existing approaches are \textit{corrective}, primarily mitigating the optimization consequences of non-IID data without changing the underlying communication interface.

\textbf{Limitations and our distinction.} FedPix does \emph{not} compete with these FL aggregation algorithms. Rather, it operates at a different layer of the federated system stack. Traditional FL addresses the question ``how should models be coordinated across clients?''; FedPix addresses the prior question ``what information should leave the client?'' The two layers are orthogonal and complementary: one could deploy a FedPix compressor at the data interface of any FL system, upload compressed representations instead of raw data, and still apply FedAvg, SCAFFOLD, or any other aggregation protocol downstream. FedPix's centralized training on a shared compressor's outputs naturally avoids the non-IID aggregation problem, but this is a consequence of its data-interface design, not a claim to replace FL aggregation.

\subsection{Federated Data Condensation}

Federated Data Condensation (FDC) reduces communication overhead by compressing raw data into small synthetic datasets. FedDM is the first to introduce data condensation into federated learning, achieving communication-efficient model updates through iterative distribution matching~\citep{xiong2023feddm}. To address the multi-round communication bottleneck, FedSD2C adopts a one-round synthetic distiller-distillate communication framework~\citep{zhang2024fedsd2c}, while CollabDM captures global data distribution through collaborative data distillation~\citep{holland2024collabdm}. \citet{zhang2024privacy} further integrated differential privacy into condensation to provide quantifiable privacy guarantees. FedDC proposes discrepancy distribution matching and collaborative knowledge distillation for IoT scenarios~\citep{ma2025communication}, and FedPAD adopts an aggregation-free prototype-based adaptive distillation paradigm~\citep{zhong2025fedpad}.

\textbf{Limitations and our distinction.} While data condensation reduces communication by compressing the \textit{number} of samples, it places substantial optimization burden on the client side and targets high-fidelity reconstruction of the original data distribution---both of which are misaligned with edge-device deployment and privacy preservation. FedPix takes a fundamentally different path: it compresses the \textit{per-sample information content} ($|z_i| \ll |x_i|$) rather than the sample count, offloads all compressor training to the server, and explicitly optimizes for task utility with a reconstruction-suppression objective. Moreover, FedPix addresses the data-interface layer (what clients upload) rather than the communication-efficiency layer (how much is uploaded). The two approaches are complementary: one could apply FedPix's per-sample compression \emph{and then} condense the resulting compressed dataset, or vice versa.

\subsection{Split Federated Learning}

Split Federated Learning (SFL) reduces client computation by transmitting only intermediate features instead of raw data~\citep{thapa2022splitfed}. \citet{turina2020combining} highlighted its efficiency-privacy balance. FLEX-SFL further improved training efficiency in heterogeneous edge environments through adaptive splitting and client selection mechanisms~\citep{flexsfl2025}. However, SFL remains vulnerable to various attacks: security analysis revealed multiple attack surfaces~\citep{khan2022security}; Unsplit demonstrated feature inversion, model stealing, and label inference~\citep{erdogan2022unsplit}; subsequent attacks include GRAMSSAT (label inference)~\citep{zhang2025gramssat}, PCAT (pseudo-client attack)~\citep{gao2023pcat}, and PPoison (poisoning attack)~\citep{liu2026ppoison}. These studies collectively expose the fragility of intermediate representation management in existing SFL methods and underscore the need for stronger privacy protections.

\textbf{Limitations and our distinction.} A critical vulnerability of SFL is that the transmitted intermediate features are not interpretable by end users, making direct human inspection of uploaded information difficult. Moreover, as prior attacks (Unsplit~\citep{erdogan2022unsplit}, GRAMSSAT~\citep{zhang2025gramssat}, PCAT~\citep{gao2023pcat}, PPoison~\citep{liu2026ppoison}) have shown, these features carry enough structural information to enable feature inversion, model stealing, and label inference. FedPix addresses this gap by transmitting compressed images that can be upsampled for inspection that preserve a visual correspondence to the original input, allowing users to inspect what is being uploaded. At the same time, our task-directed compression actively suppresses reconstructability, providing stronger reconstruction resistance than the passive feature exposure in SFL.

\subsection{Differential Privacy Mechanisms}

Differential Privacy (DP) provides quantifiable privacy guarantees by injecting calibrated noise into data or model parameters, and has been widely adopted in federated learning frameworks to defend against privacy attacks. DP-Pix first combined differential privacy with image pixelization, achieving provable privacy guarantees by injecting noise into the pixelization process~\citep{dppix2019}; DP-GLOW, based on invertible generative models, adds noise in the latent space to achieve privacy protection for generated images~\citep{dpvae2023}. LDP-Slicing proposed a local differential privacy perturbation mechanism based on randomized bit-plane slicing, which reduces noise variance by processing data plane by plane, thereby improving data utility while preserving privacy~\citep{ldpslicing2026}. Furthermore, DP-FedAvg combines differential privacy with the federated averaging algorithm, enabling privacy budget composition across multiple training rounds through privacy accounting~\citep{mcmahan2017learning}; PRISM-FL proposes an obfuscated image synthesis framework under client-level differential privacy, balancing privacy and utility by generating datasets that resemble the private data distribution yet exhibit substantial visual differences from the original data~\citep{soufleri2025prism}.

\textbf{Limitations and our distinction.} The fundamental drawback of DP-based data perturbation is its \textit{indiscriminate} nature: noise is added to all information dimensions regardless of their relevance to the downstream task, leading to a severe utility-privacy trade-off that worsens with data dimensionality. FedPix overcomes this by replacing blind noise with a \textit{learnable, task-directed utility-privacy filter}. The compressor is trained on the server to selectively preserve task-relevant cues while suppressing those that enable reconstruction, all without accessing raw client data. Our experiments compare empirical utility and attack metrics for selected configurations; they do not establish superiority at a matched formal privacy budget.

\subsection{Lossy Compression and Learned Compression}

Lossy compression has long been studied as a means to reduce storage and transmission costs. Traditional codecs such as JPEG~\citep{wallace1992jpeg}, JPEG2000, and HEVC~\citep{sullivan2012overview} rely on hand-crafted transforms (e.g., DCT, wavelet) and quantization to achieve compression. While these methods are computationally efficient on edge devices and produce human-readable images, they are \textit{non-learnable} and optimize for reconstruction fidelity rather than task utility or privacy. Consequently, they cannot adapt to downstream tasks and provide no formal protection against reconstruction attacks.

More recently, learned image compression (LIC) has emerged as a powerful alternative, employing neural autoencoders to optimize the rate-distortion trade-off end-to-end~\citep{balle2018variational,balle2017end,minnen2018joint}. These methods typically encode images into compact latent representations that are entropy-coded into bitstreams. While they achieve superior compression ratios, they suffer from two key limitations in the context of privacy-aware federated learning: (1) the transmitted latent representations or bitstreams are \textit{not human-readable}, violating user auditability (\textbf{R1}); and (2) the encoder often requires significant computation or even task-specific fine-tuning on the client side, conflicting with edge-device friendliness (\textbf{R2}). Furthermore, like traditional codecs, LIC methods optimize for reconstruction quality, which inherently conflicts with the privacy goal of non-reconstructability (\textbf{R3})~\citep{blau2018perception,blau2019rethinking}. The seminal work of Blau and Michaeli~\citep{blau2019rethinking} formalized the rate-distortion-perception trade-off, showing that perfect perceptual quality (i.e., indistinguishability from the original) is fundamentally at odds with low distortion---a tension that directly translates to the privacy domain, where high reconstruction fidelity implies high privacy leakage.

A few recent works have explored privacy-oriented compression with fixed (non-learned) transformations. DeepFixCX~\citep{gaudio2023deepfixcx} proposes an explainable and privacy-oriented medical image compression framework based on wavelet packet transforms and spatial pooling. By removing local spatial information and suppressing edge direction via $\ell_1$ pooling, DeepFixCX generates human-readable compressed images that mitigate patient re-identification while maintaining diagnostic utility. However, DeepFixCX is \textit{non-learnable}: its wavelet filters and pooling operations have fixed weights, meaning the compression strategy cannot adapt to new tasks or data distributions beyond the medical imaging scenarios for which it was designed. Moreover, while the compression is fast (650--1700 images per second on a laptop CPU/GPU), the client still bears the full computational burden of the transform and pooling, and the method does not explicitly optimize against reconstruction attacks such as model inversion. These limitations highlight the need for a \textit{learnable} privacy-aware compression framework that shifts training overhead to the server while keeping client-side inference minimal.

Other works have explored task-oriented or semantic compression~\citep{liu2019classification,wang2025task}, which prioritizes task utility over pixel-level fidelity. However, these approaches still typically require client-side training or transmit non-auditable features. For instance, \citet{liu2019classification} analyzed the classification-distortion-perception trade-off, showing that optimizing for classification accuracy does not automatically guarantee privacy against reconstruction. Similarly, \citet{wang2025task} proposed task-oriented lossy compression frameworks under data, perception, and classification constraints, but the training burden remains on the client side.

\textbf{Limitations and our distinction.} To the best of our knowledge, FedPix is the first framework that provides reconstruction resistance through a capacity-constrained representation interface while simultaneously satisfying all three requirements: it transmits human-readable compressed images, shifts all training overhead to the server, and explicitly optimizes for dissimilarity and non-reconstructability.

\subsection{Task-Oriented Image Compression}

Task-oriented image compression has emerged as a paradigm shift from traditional rate-distortion optimization toward preserving information relevant to specific downstream tasks such as classification, detection, or segmentation~\citep{liu2019classification,wang2025task}. Unlike conventional codecs that minimize pixel-level distortion for human perception, these methods integrate task-specific objectives---e.g., classification cross-entropy or segmentation mIoU---into the compression pipeline, enabling extreme bitrate reduction with minimal utility degradation~\citep{Codevilla2021,KawawaBeaudan2022}. A prominent line of work organizes the latent representation according to task-specific mutual information, selectively retaining features that are informative for analytics while discarding those that are not~\citep{alvar2021scalable}. Azizian and Baji{\'c}~\citep{azizian2024ppa} propose a privacy-preserving autoencoder (PPA) that uses adversarial training to force the encoder to discard pre-specified private attributes (e.g., faces, license plates) from compressed representations while preserving task-relevant features for collaborative object detection. The encoder is trained on a public dataset at the server and then deployed to edge devices for single-pass inference. More recent efforts formulate the problem as rate-distortion-utility joint optimization, where the utility loss directly guides the compressor to produce task-favorable representations for downstream analytics such as classification, segmentation, and detection~\citep{Kubiak2021,Ge2024}. InfoShape further leverages mutual information neural estimators to train a lossy encoder that explicitly trades off privacy and utility via information-theoretic criteria~\citep{Esfahanizadeh2023}.

However, these methods share a fundamental assumption: the compressor can directly access original data during optimization. They mainly optimize compression efficiency for downstream analytics, rather than designing a privacy-controlled interface for federated data sharing. In a federated setting where raw data never leaves the client, this assumption breaks down, and the optimization objective must be reformulated to operate without direct data access.

\textbf{Limitations and our distinction.} These methods either (i) operate on latent features that are neither human-readable nor auditable~\citep{alvar2021scalable,azizian2024ppa}, (ii) require clients to run computationally heavy encoders or even perform local training~\citep{Esfahanizadeh2023}, or (iii) treat privacy protection as a secondary concern rather than a first-class objective~\citep{Codevilla2021}. FedPix addresses these gaps by introducing a Federated Data Interface in which the encoder is optimized on the server via an ``upsample-recompress'' closed loop without ever accessing raw client data, clients perform only a single lightweight forward pass, and the output remains a human-readable compressed image that users can audit. To the best of our knowledge, FedPix is the first framework that realizes a federated data interface with explicit privacy protection, task-aware representation learning, and server-side optimization without raw data access in a single cohesive design.

\subsection{Positioning of Our Work}

FedPix occupies a distinct position in the federated ecosystem. It is not a model-aggregation FL algorithm, a data-condensation method, a split-learning variant, or a DP mechanism---though it shares motivations with each. FedPix is best understood as a \textbf{Federated Data Interface}: a principled mechanism for clients to share task-useful yet privacy-limited representations with a server. It simultaneously ensures user auditability (human-readable compressed images), edge-device friendliness (single forward pass on client), and a balanced trade-off between dissimilarity, non-reconstructability, and task utility. By decoupling encoder training from client-side inference and leveraging upsampled compressed data as optimization proxies, FedPix satisfies all three requirements without client-side optimization or parameter aggregation. The shared encoder, trained centrally and distributed periodically, naturally circumvents the client drift issue common under non-IID data distributions. We emphasize that FedPix is orthogonal and complementary to existing FL methods: it addresses the \emph{data interface} layer (what leaves the client), while traditional FL addresses the \emph{model coordination} layer (how models are aggregated).

To complement the narrative survey above, Table~\ref{tab:comparison} provides a structured comparison of representative methods along the key dimensions identified in the Introduction: reconstruction resistance, underlying mechanism, limitations of existing approaches, our innovations, and edge-device friendliness.

\begin{table*}[tbp]
\centering
\caption{Comparison of representative methods---FedAvg'17~\citep{mcmahan2017communication}, SCAFFOLD'20~\citep{karimireddy2020scaffold}, FedProx'20~\citep{li2020fedprox}, PARS'26~\citep{pars2026}, FedDM'23~\citep{xiong2023feddm}, CollabDM'24~\citep{holland2024collabdm}, FedSD2C'24~\citep{zhang2024fedsd2c}, SFL'20~\citep{thapa2022splitfed}, FLEX-SFL'25~\citep{flexsfl2025}, DP-Pix'19~\citep{dppix2019}, DP-GLOW'23~\citep{dpvae2023}, LDP-Slicing'26~\citep{ldpslicing2026}, Scalable Privacy'21~\citep{alvar2021scalable}, PPA'24~\citep{azizian2024ppa}, JPEG'92~\citep{wallace1992jpeg}, and DeepFixCX'23~\citep{gaudio2023deepfixcx}---and FedPix. Columns summarize privacy mechanisms, limitations, the FedPix design, and client efficiency.}
\label{tab:comparison}
\small
\setlength{\tabcolsep}{3pt}
\renewcommand{\arraystretch}{1.5}

\setlength{\heavyrulewidth}{1.5pt}
\setlength{\lightrulewidth}{0.8pt}

\resizebox{\textwidth}{!}{%
\begin{tabular}{>{\raggedright\arraybackslash}p{2.0cm}|p{1.7cm}|p{4.0cm}|p{2.8cm}|p{4.8cm}|>{\raggedright\arraybackslash}p{1.3cm}}
\toprule
\textbf{Method} &
\textbf{Privacy Aware?} &
\textbf{Protection Mechanism} &
\textbf{Limitations} &
\textbf{Our Innovation} &
\textbf{Edge-friendly?} \\
\midrule

\multicolumn{6}{l}{\textbf{Traditional FL}} \\
\midrule

FedAvg '17 & $\checkmark$ & \multirow{4}{4.0cm}{Uploads model parameters instead of raw data} & \multirow{4}{2.8cm}{Uploaded data is not user-auditable} & \multirow{4}{4.8cm}{Uploads human-interpretable compressed images instead of model parameters} & $\times$ \\
SCAFFOLD '20 & $\checkmark$ & & & & $\times$ \\
FedProx '20 & $\checkmark$ & & & & $\times$ \\
PARS '26 & $\checkmark$ & & & & $\times$ \\
\midrule

\multicolumn{6}{l}{\textbf{Federated Data Condensation}} \\
\midrule

FedDM '23 & $\checkmark$ & \multirow{3}{4.0cm}{Uploads synthetic data (or its encoding) templated on real data instead of raw data} & \multirow{3}{2.8cm}{Optimizes for high fidelity; synthetic data resembles original data} & \multirow{3}{4.8cm}{Optimizes for privacy-utility information filtering rather than high fidelity} & $\times$ \\
CollabDM '24 & $\checkmark$ & & & & $\checkmark$ \\
FedSD2C '24 & $\checkmark$ & & & & $\times$ \\
\midrule

\multicolumn{6}{l}{\textbf{Split Federated Learning}} \\
\midrule

SFL '20 & $\checkmark$ & \multirow{2}{4.0cm}{Uploads intermediate features instead of raw data} & \multirow{2}{2.8cm}{Uploaded data is not user-auditable} & \multirow{2}{4.8cm}{Uploads user-auditable compressed images instead of intermediate features} & $\checkmark$ \\
FLEX-SFL '25 & $\checkmark$ & & & & $\checkmark$ \\
\midrule

\multicolumn{6}{l}{\textbf{Differential Privacy for Data Rather Than for the Model}} \\
\midrule

DP-Pix '19 & $\checkmark$ & \multirow{3}{4.0cm}{Uploads noise-perturbed data (or its encoding) instead of raw data} & \multirow{3}{2.8cm}{Blind noise addition sacrifices utility for privacy} & \multirow{3}{4.8cm}{Optimizes for privacy-utility information filtering rather than blindly erasing both with equal noise} & $\checkmark$ \\
DP-GLOW '23 & $\checkmark$ & & & & $\checkmark$ \\
LDP-Slicing '26 & $\checkmark$ & & & & $\checkmark$ \\
\midrule

\multicolumn{6}{l}{\textbf{Task-Oriented Compression}} \\
\midrule

Scalable Privacy '21 & $\checkmark$ & \parbox{4.0cm}{\raggedright Uploads latent features instead of raw data, with privacy-prioritized feature sorting and selective uploading} & \parbox{2.8cm}{\raggedright Optimizes feature ranking rather than the encoder; client bears main computational cost; not a federated data interface} & \parbox{4.8cm}{\raggedright Learns a task-aware privacy--utility encoder rather than ranking fixed features; server bears main computational cost; designed as a federated data interface} & $\times$ \\
PPA '24 & $\checkmark$ & \parbox{4.0cm}{\raggedright Uses adversarial training to force the encoder to discard private attributes (e.g., faces, license plates) from compressed representations while preserving task-relevant features for object detection. The encoder is trained on a public dataset at the server, then deployed to edge devices for single-pass inference.} & \parbox{2.8cm}{\raggedright Outputs unreadable latent features that users cannot audit. Requires pre-specifying privacy attributes to protect (e.g., faces, license plates); retraining is needed for different privacy targets.} & \parbox{4.8cm}{\raggedright Client uploads human-readable compressed images that support user auditing; no need to pre-specify privacy attributes---the capacity bottleneck adaptively filters information; designed as an iterative federated data interface rather than a one-shot deployable filter.} & $\checkmark$ \\
\midrule

\multicolumn{6}{l}{\textbf{Lossy Compression}} \\
\midrule

JPEG '92 & $\times$ & -- & -- & Optimizes for privacy-utility filtering rather than high fidelity & $\checkmark$ \\
DeepFixCX '23 & $\checkmark$ & \parbox{4.0cm}{\raggedright 1.~Multiple privacy operators;\\ 2.~Multi-pixel averaging (equivalent to downscaling)} & \parbox{2.8cm}{\raggedright Blind noise addition sacrifices utility for privacy} & \parbox{4.8cm}{\raggedright Optimizes for privacy-utility information filtering rather than blindly erasing both with equal noise} & $\checkmark$ \\
\midrule

\rowcolor{gray!20}
\textbf{FedPix (Ours)} & $\checkmark$ & \parbox{4.0cm}{\raggedright 1.~Learns a task-aware representation shared by all clients;\\ 2.~Server-side closed-loop optimization via upsample-recompress;\\ 3.~Spatial downsampling limits representation size; leakage is evaluated empirically} & -- & -- & $\checkmark$ \\
\bottomrule
\end{tabular}%
}
\end{table*}

\end{document}